\documentclass[a4paper,11pt,superscriptaddress,preprintnumbers]{article}

\usepackage{jcappub} 

\usepackage{amssymb,amsmath,latexsym,bm,amsfonts}

\usepackage{longtable}
\usepackage{color,xcolor}
\usepackage{indentfirst}
\usepackage[margin=1in]{geometry}
\usepackage{mathtools,tensor,graphicx,}

\usepackage{stmaryrd}

\usepackage{tikz,colortbl}
\usetikzlibrary{calc}
\usepackage{zref-savepos}
\usepackage{array}

\newcommand{\vp}{\varphi}
\newcommand{\Gc}{\mathcal{G}}
\newcommand{\Cc}{\mathcal{C}}
\newcommand{\p}{\partial}
\newcommand{\gmn}{g_{\mu\nu}}

\numberwithin{equation}{section}

 \usepackage{cleveref}
 
\title{\boldmath Multi-scalar theories of gravity with direct matter couplings and their parametrized post-Newtonian parameters}
\author[a,b,c]{Osmin Lacombe}
\author[a,d]{Shinji Mukohyama}

\affiliation[a]{Center for Gravitational Physics and Quantum Information, Yukawa Institute for Theoretical Physics, Kyoto University, Sakyo-ku, Kyoto 606-8502, Japan}
\affiliation[b]{Dipartimento di Fisica e Astronomia, Universit\`a di Bologna, via Irnerio 46, 40126, Bologna, Italy}
\affiliation[c]{INFN, Sezione di Bologna, viale Berti Pichat 6/2, 40127 Bologna, Italy}
\affiliation[d]{Kavli Institute for the Physics and Mathematics of the Universe, The University of Tokyo Institutes for Advanced Study, The University of Tokyo, Kashiwa, Chiba, 277-8583, Japan}

\emailAdd{deriusosmin.lacombe@unibo.it}
\emailAdd{shinji.mukohyama@yukawa.kyoto-u.ac.jp}

\abstract{We study theories of gravity including, in addition to the metric, several scalar fields in the gravitational sector. The particularity of this work is that we allow for direct couplings between these gravitating scalars and the matter sector, which can generally be different for the source and the probe of gravity, in addition to the universal interactions generated by the Jordan frame metric. The weak gravity regime of this theory, which would describe solar-system experiments, is studied using the parametrized post-Newtonian (PPN) formalism. We derive the expression of the ten parameters of this formalism. Among them, $\zeta_3$ and $\zeta_4$ are modified with respect to their values in the theories without direct couplings. This fact holds even after eliminating the direct couplings between the gravitating scalars and the energy density of the source, by redefinition of the Jordan frame. All other PPN parameters are insensitive to the direct couplings once in the correctly identified Jordan frame. When direct couplings are different for the source and the probe of gravity, they make non-relativistic probes deviate from the geodesics of the PPN metric in this frame, already at Newtonian order. Such couplings would thus be directly detectable and would have excluded by experiments. This shows that, contrary to the claims in the recent literature, it is impossible to screen the presence of gravitating scalars relying only on a curved target space and direct couplings to matter.} 

\begin{document}
\begin{flushright} { YITP-23-10\\IPMU23-0004}  \end{flushright}
\vspace{-1.4cm}

\maketitle
\flushbottom

\section{Introduction}\label{introduction}

Massless or very light gravitating scalar fields naturally arise in various cosmological models motivated by high-energy constructions. In theories with extra dimensions, such as string theory, they are naturally related to the geometry of the extra-dimensional space and are hence generally called geometric moduli, radions or yet dilatons. The historical prototypical example of a theory coupling a dilaton to gravity is the Brans-Dicke theory \cite{Brans:1961sx}. In effective theories descending from string theory, these scalar fields are usually associated with scale-invariant structures at the leading order in perturbation theory \cite{Witten:1985xb,Nilles:1986cy,Burgess:1985zz,Ellis:1983sf,Burgess:2020qsc} and are thus massless at this level. Upon mechanisms breaking their no-scale structures, such as the inclusion of quantum corrections, the scalars can acquire small masses stabilizing them to certain vacuum expectation values. If such stabilization indeed happened in the far past history of our Universe, at high energy scales, these scalars are not active anymore and should just be included in the effective vacuum energy density. 

On another side, scalars that remained massless or extremely light today, hence not stabilised and cosmologically active, can have numerous applications in modified theories of gravity used to construct dark energy or dark matter models. For instance, they can lead to equivalent descriptions of higher-order theories of gravity \cite{Gottlober:1989ww} or be quintessence candidates \cite{Ratra:1987rm,Caldwell:1997ii,Copeland:1997et,Tsujikawa:2013fta,Cicoli:2018kdo}. However, cosmologically active (almost-)massless scalars would necessarily mediate fifth forces for matter and if the scalars are universally coupled to matter, these forces would typically be of gravitational strength \cite{Jordan:1949zz}. These effects would then be accessible in weak-gravity regions and observable in experiments constraining this regime, such as solar-system experiments \cite{Will:2014kxa}. The latter indeed highly constrain metric theories of gravity and show that deviations from general relativity have to be extremely small. This in turns highly constrains the strength of universal couplings from gravitating scalars to matter, ruling out their cosmological interest unless some screening mechanism takes place.

Numerous modifications of gravity prove successful at cosmological scales but dangerously change physics at shorter length scales, in the Newtonian regime. Such theories cannot thus be reasonably considered valid at any scale. One thus usually invokes screening mechanisms hiding their features, such as the presence of very light scalars, in solar-system experiments. Such mechanisms rely on non-linearities in the potential \cite{Khoury:2003aq}, couplings \cite{Damour:1994zq,Olive:2007aj}  or kinetic terms \cite{Vainshtein:1972sx} of the scalar fields. See \cite{Brax:2013ida} for a review. 

In recent papers \cite{Burgess:2021qti,Brax:2022vlf}, the authors have studied the possibility of a mechanism that could potentially hide the presence of cosmologically active scalars in the weak-field quasi-static regime. The authors rely on direct couplings of scalar fields to matter fields, in addition to the standard universal coupling through the metric. As such couplings violate the equivalence principle, they should be small enough to remain undetectable by experiments testing this principle on Earth. The studies mentioned above argue that even for very small couplings, the weak gravity regime can be modified so to screen the presence of the very light gravitating scalars. The authors have focused particularly on a model containing an axion and a dilaton, with direct coupling between matter and the axion. 

Motivated by this work, we investigate the weak gravity regime of general multi-scalar theories with the inclusion of direct couplings in addition to the universal ones induced by the Jordan frame metric. We thus study scalar-tensor theories including several massless gravitating scalars, with direct matter couplings. The gravity sector, containing the metric and gravitating scalars with curved scalar target space but no scalar potential, will be coupled to matter through the Jordan frame metric and additional couplings. To study this theory in the weak-field quasi-static regime, we will make use of the parametrized post-Newtonian (PPN) formalism, which is a natural framework to compare theories of gravity in this regime and has been developed gradually throughout the last century building on the early work of Eddington \cite{1923mtr..book.....E,1962saa..conf..228R,1967rta1.book..105S,Will:1971zzb,Will:1972zz,Will:1973zz}.

The rest of the paper is organised as follows. In \Cref{section:lagrangianandequations} we present the framework studied in the paper, by presenting the Lagrangian and associated equation of motions of the gravitational theory under study, both in the Jordan and Einstein frames. For concreteness, we show how the framework applies to the simple examples of the Brans-Dicke theory and the axio-dilaton theory motivating this work. We then study in \Cref{section:PPNsection} the quasi-static weak-field regime of these theories, making use of the parametrized post-Newtonian formalism. We derive the expression for the ten PPN parameters in multi-scalar theories with direct couplings between gravitating scalars and matter fields. We then apply the obtained formulae to the two aforementioned examples. Eventually, we study in \Cref{section:Observations} if direct couplings can change classical tests of theories of gravity giving access to the PPN metric, and thus the PPN parameters. We present in \Cref{section:conclusions} a summary of our results and discuss possible future directions along this work. The paper also includes \Cref{appendix:usefulrelations} which presents definitions and identities related to the target-space functions and PPN functionals used in the main body.

\section{Multi-scalar theory of gravity with direct coupling } \label{section:lagrangianandequations}

Scalar-tensor theories of gravity have been studied extensively in the literature \cite{Gottlober:1989ww,Damour:1992we}. In the present work we study the case of massless gravitating scalars, hence coupled non-minimally to gravity, in the presence of direct couplings between these scalars and the matter fields. In this section, we first present the action and set notations for the theories we consider in the rest of this work. We show how the presence of direct couplings modify field equations for the gravitational fields and support our discussion by taking two simple examples.
~\\

 \subsection{Action with direct couplings and equations of motion} 
 
 \paragraph{Jordan frame action} We study scalar-tensor theories of gravity containing $N$ several scalar fields $\varphi^a$, $a=1,\ldots, N$, with non-minimal coupling to gravity defined by a function $F$ and kinetic terms defined on a target space parametrized by a metric $\mathcal{G}_{ab}$. We thus consider the following four-dimensional Jordan frame action
\begin{equation}
S=S_g+S_m=\frac{M^2}2\int d^4x \sqrt{-g}\left [ F(\varphi^a) R - \mathcal{G}_{ab}(\vp^c)\partial_{\mu} \varphi^a\partial_{\nu} \varphi^b g^{\mu\nu} \right ]+ \int d^4x \, \mathcal{L}_m(g_{\mu\nu},\chi, \varphi^a). \label{Action}
\end{equation}
This action is written using the Jordan frame metric $g_{\mu\nu}$ and in the matter Lagrangian $\mathcal{L}_m$ matter fields are denoted by $\chi$. As motivated in \cref{introduction}, we allow for additional (weak) couplings between the gravitating scalars and matter, as seen from the $\varphi^a$  dependence  of $ \mathcal{L}_m$. The presence of these terms would violate the so-called {\it universal coupling} of matter to gravitational fields. In the above action, $M$ is a mass scale. 

The scalar equations of motion derived from this action read
\begin{equation}
\Box\varphi^a+\Gamma^a_{bc}\p\varphi^b\p\varphi^c+\frac 12 F^a  R + \frac{1}{2M^2}\Cc^a=0. \label{scalar-eom1}
\end{equation}
Partial derivatives with Latin indices are taken with respect to scalar fields, $\p_a\equiv \p/\p\vp^a$ and the metric $\Gc_{ab}$ (inverse metric $\Gc^{ab}$) is used to lower (raise) scalar target-space indices. As in the rest of the paper, we make use of the notations
\begin{equation}
F_a\equiv \p_aF=\frac{\p F}{\p\vp^a},  \qquad F_{ab}\equiv \p_{ab}F, \qquad F^b=\Gc^{ba}F_a.
\end{equation}
In \cref{scalar-eom1} we used the Christoffel symbol $\Gamma^a_{bc}$ defined as the Levi-Civita connection for the target-space metric
\begin{equation}
\Gamma^a_{bc}=\frac 12 \Gc^{ad}(\p_b\Gc_{cd}+\p_c\Gc_{bd}-\p_d\Gc_{bc}),
\end{equation}
and omitted space-time summation in the scalar kinetic terms in \cref{scalar-eom1}, as will be done in the rest of the paper. One should thus read as usual $\p\vp^b\p\vp^c\equiv \p^{\sigma}\vp^b\p_{\sigma}\vp^c$. We have also introduced the matter-scalar coupling functions, which will be the main ingredients of the present work and are defined through
\begin{equation}
\Cc_a\equiv\frac{2}{\sqrt{-g}}\frac{\delta{S}_m}{\delta{\vp^a}}. \label{def-Ca}
\end{equation}

The Einstein equations derived from our action \eqref{Action} read
\begin{align}
F\left(R_{\mu\nu}-\frac 12 g_{\mu\nu}R\right)= &\frac 1{M^2} T_{\mu\nu} + \Gc_{ab}\left(\p_{\mu}\vp^a\p_{\nu}\vp^b-\frac 12 g_{\mu\nu} \, \p \vp^a \p\vp^b\right) \nonumber \\
&+ F_a\left(\nabla_{\mu}\p_{\nu}\vp^a-g_{\mu\nu}\Box \vp^a\right)+F_{ab}\left(\p_{\mu}\vp^a\p_{\nu}\vp^b-g_{\mu\nu}\p\vp^a \p\vp^b\right). \label{metric-eom1}
\end{align}
We introduced the usual stress-energy tensor
\begin{equation}
T^{\mu\nu}=\frac2{\sqrt{-g}}\frac{\delta S_m}{\delta{g_{\mu\nu}}}, \quad T=T^{\mu\nu}g_{\mu\nu}. \label{def-T}
\end{equation}
Taking the trace of \cref{metric-eom1} leads to the following expression for the Ricci scalar
\begin{equation}
R=-\frac{1}{F M^2} T+\frac{1}{F}\left(\Gc_{ab}+3F_{ab}\right)\p\vp^a\p\vp^b+3\frac{F_a}{F}\Box\vp^a, 
\end{equation}
which can be used in \cref{scalar-eom1,metric-eom1} to express the scalar equations and trace-reversed Einstein equations as
\begin{align}
&\Box\varphi^a+\frac 3{2} \frac{F^a}{F}{F_b} \,\Box \vp^b+\left[\Gamma^a_{bc}+\frac{F^a}{2F}(\Gc_{bc}+3F_{bc})\right]\p\varphi^b\p\varphi^c = \frac1{2M^2}\Big(\frac{F^a}{F} T -\Cc^a\Big),&\label{scalar-eom2}\\
&F R_{\mu\nu}-F_a \Big(\nabla_{\mu}\p_{\nu}\vp^a+\frac 12 g_{\mu\nu} \Box\vp^a\Big)\hspace{3cm}&\nonumber\\
&\hspace{2cm}-\frac 12 \gmn F_{ab} \p\vp^a\p\vp^b - \left(\Gc_{ab}+F_{ab}\right)\p_{\mu}\vp^a\p_{\nu}\vp^b =  \frac1{M^2}\Big(T_{\mu\nu}-\frac 12 \gmn T\Big).& \label{metric-eom2}
\end{align}
We recall that $F$, $F_a$, $F_{ab}$, $\Gc_{ab}$, $\Gamma^a_{bc}$ are  target-space functions depending on the scalar fields $\vp^a$.

\paragraph{Einstein frame action} One could also study the same theory in the Einstein frame, where there is no prefactor in front of the Ricci scalar. Although the PPN formalism is expressed in the Jordan frame, some theories are naturally obtained and easily interpreted in a frame with the canonical Einstein-Hilbert term. As this is the case for the axio-dilaton example we study later, we now detail the explicit relation between the two formulations. They are equivalent to each other and can be related by using the Einstein frame metric $g^E$ defined through 
\begin{equation}
g_{\mu\nu}=\frac{1}{F(\varphi^a)}g^E_{\mu\nu}.\label{JFtoEF}
\end{equation}
The action \eqref{Action} can thus be written in the Einstein frame as
\begin{equation}
S=\frac{M^2}2\int d^4x \sqrt{-g^E}\left [ R^E - \mathcal{G}^E_{ab}\partial_{\mu} \varphi^a\partial_{\nu} \varphi^b g^{E\,\mu\nu} \right ]+ \int d^4x \, \mathcal{L}_m\Big(F^{-1}g^E_{\mu\nu},\chi, \varphi^a\Big). \label{ActionEF}
\end{equation}
The Einstein frame target-space metric $\mathcal{G}^E_{ab}$ now includes new contributions coming from the Weyl rescaling of the Ricci scalar, which schematically reads $R=F(\varphi^a)R^E+f(\varphi^a,\partial \varphi^a)$. The relation between the two target-space metrics can be derived by expressing exactly the additional terms of the Lagrangian appearing due to $f(\varphi^a,\partial \varphi^a)$. Indeed, we shall have, up to total derivatives:
 \begin{align}
 \sqrt{-g}\left(F R +  \mathcal{G}_{ab} \partial_{\mu} \varphi^a\partial_{\nu} \varphi^b g^{\mu\nu}\right)=\sqrt{-g^E}\left(R^E+  \mathcal{G}^E_{ab} \partial_{\mu} \varphi^a\partial_{\nu} \varphi^b g^{E\,\mu\nu}\right),
 \end{align}
 which, using \cref{JFtoEF} leads to
 \begin{align}
\Big(F\mathcal{G}^E_{ab}- \mathcal{G}_{ab}\Big) g^{E\,\mu\nu} \partial_{\mu} \varphi^a\partial_{\nu} \varphi^b &=F R^E-R =\frac{3}{2}\frac{F_aF_b}{F}\partial_{\mu} \varphi^a\partial_{\nu} \varphi^b   g^{E\,\mu\nu}.
\end{align}
The last equality holds in $d=4$ dimensions and up to total derivatives. This leads to the simple relation:
\begin{equation}
\mathcal{G}^E_{ab}=\frac{1}{F}\,\mathcal{G}_{ab}+\frac{3}{2}\frac{F_aF_b}{F^2}. \label{relationTSmetrics}
\end{equation}
We again stress that \cref{ActionEF} is obtained after partial integration. The stress-energy tensor $T^E_{\mu\nu}$ is defined as in \cref{def-T}, obtained by varying the matter action with respect to the Einstein frame metric, reads
\begin{equation}
T^{E\, \mu\nu}=\frac2{\sqrt{-g^E}}\frac{\delta S_m}{\delta{g^E_{\mu\nu}}}=\frac{1}{F^3}T^{\mu\nu}, \quad T^E_{\mu\nu}=\frac{1}{F}T_{\mu\nu}, \quad T^E=g^E_{\mu\nu}T^{E\, \mu\nu}=\frac{1}{F^2} T.  \label{TE}
\end{equation}
 The Einstein frame matter-scalar coupling functions $\Cc^{E\,a}$ also differ from the ones in the Jordan frame, not only due to the $\sqrt{-g}$ factor in their definition \cref{def-Ca}, but also due to the fact that the universal coupling contains a term in the scalar fields. They indeed read
\begin{align}
\Cc^{E}_a&=\frac{2}{\sqrt{-g_E}}\frac{\delta{S}_m(F^{-1}g^E_{\mu\nu},\chi,\vp^a)}{\delta{\vp^a}}= \frac{\sqrt{-g}}{\sqrt{-g_E}}\mathcal{C}_a+\frac{2}{\sqrt{-g_E}}\frac{\delta S_m}{\delta g^E_{\mu\nu}}\frac{\delta g^E_{\mu\nu}}{\delta g_{\mu\nu}}\frac{\delta g_{\mu\nu}}{\delta \varphi^a}\nonumber\\
&=\frac{\sqrt{-g}}{\sqrt{-g_E}}\mathcal{C}_a-\frac{F_a}{F}T^E=\frac{1}{F^2}\left(\mathcal{C}_a-\frac{F_a}{F}T\right). \label{CcE}
\end{align}
The equations of motion in the Einstein frame can be written in terms of these new quantities as
\begin{align}
&\Box^E\varphi^a+\Gamma^{E\,a}_{\,\,\,\,\,bc}\p\varphi^b\p\varphi^c = - \frac1{2M^2} \Cc^{E\, a},&\label{scalar-eom2EF}\\
&R^E_{\mu\nu}-\Gc^E_{ab}\p_{\mu}\vp^a\p_{\nu}\vp^b =  \frac1{M^2}\Big(T^E_{\mu\nu}-\frac 12 \gmn T^E\Big).& \label{metric-eom2EF}
\end{align}
They are simply derived from the Einstein frame action \eqref{ActionEF} but can also be induced from \cref{scalar-eom2,metric-eom2} making use of relations \eqref{relationTSmetrics} to \eqref{CcE}. Although these field equations seem simpler than the Jordan frame ones, PPN parameters are computed in this latter frame. This is the case because the Jordan frame metric appears in matter field kinetic terms so that particles follow geodesics of the Jordan frame metric. 

\subsection{Particular cases}\label{subsection:particularcases}

We now apply the above generic notations for multi-scalar gravity to two particular examples, the Brans-Dicke(-like) theory and the axio-dilaton theory. As it contains one single scalar field, the first one is not even a multi-scalar but is nevertheless useful for the rest of the discussion.

\paragraph{Brans-Dicke scalar-tensor theory} We start with the study of the simple Brans-Dicke theory \cite{Brans:1961sx}, which contains only one scalar field with kinetic term parametrized by $\omega$. Even though in the original version of the theory a constant $\omega$ was studied, we relax this condition as in the extended case studied by \cite{Bergmann:1968ve,Wagoner:1970vr,Nordtvedt:1970uv}. The theory is thus described by the Jordan frame action
\begin{equation}
S_{BD}=\frac{M^2}2\int d^4x \sqrt{-g}\Big [\phi R -\frac{\omega(\phi)}{\phi}(\p\phi)^2\Big ] + \int d^4x \, \mathcal{L}_m(g_{\mu\nu},\chi). \label{BDAction}
\end{equation}
This action can be seen as a particular case of \cref{Action} by taking schematically 
\begin{align}
\varphi^a=\phi, \quad \Gc_{ab}\p\vp^a\p\vp^b=\frac{\omega(\phi)}{\phi}(\p\phi)^2, \quad F(\vp^a)=\phi, \quad\Cc_a=0.
\end{align}
These definitions lead to the following target-space functions
\begin{equation}
 \quad F_a\equiv\p_aF=1, \,\, F_{ab}\equiv\p_{ab}F=0, \quad F^a=\Gc^{ab}{F}_b=\frac\phi \omega, \quad \Gamma^a_{bc}=\frac{1}2\frac{\phi}{\omega} \p_{\phi}\Big(\frac w\phi\Big),
\end{equation}
which, according to \cref{scalar-eom2,metric-eom2}, give the equations of motion
\begin{align}
&(3+2\omega)\Box\phi+\frac 1{2\omega}\frac{d\omega}{d\phi}(\p\phi)^2=\frac T{M^2},\label{scalar-eomBDJ}\\
&\phi R_{\mu\nu}-\nabla_{\mu}\p_{\nu} \phi - \frac12g_{\mu\nu}\Box\phi- \frac w{\phi}\p_{\mu}\phi\p_{\nu}\phi= \frac1{M^2}\Big(T_{\mu\nu}-\frac 12 \gmn T\Big).\label{metric-eomBDJ}
\end{align}
The Brans-Dicke theory can also be expressed in the Einstein frame through the Weyl rescaling 
\begin{equation}
g_{\mu\nu}=\frac{1}{F(\phi)}g^E_{\mu\nu}=\frac 1\phi  \, g^E_{\mu\nu},
\end{equation}
which allows to rewrite the action \eqref{BDAction} as
\begin{equation}
S_{BD}=\frac{M^2}2\int d^4x \sqrt{-g^E}\Big [ R^E -\Big(\frac 32 + w(\phi)\Big)\frac 1{\phi^2}(\p\phi)^2\Big] + \int d^4x \, \mathcal{L}_m\Big(\frac1\phi g^E_{\mu\nu},\chi\Big). \label{BDActionE1}
\end{equation}
In the case of constant $w(\phi)=w$, one obtains the canonical scalar kinetic term by defining a new variable $\phi^E$ so that
\begin{equation}
(\p\phi^E)^2=\Big(\frac 32 + \omega\Big)\frac 1{\phi^2}(\p\phi)^2,
\end{equation}
and thus
\begin{equation}
 \phi=\exp\Big(\pm \sqrt{\frac2{3+2\omega}}\phi^E\Big)=F. \label{JtoEscalarBD}
\end{equation}
The action for this canonical scalar in the Einstein frame hence reduces to
\begin{equation}
S_{BD}=\frac{M^2}2\int d^4x \sqrt{-g^E}\Big [ R^E -(\p\phi^E)^2 \Big]+ \mathcal{L}_m\Big(\frac{1}{F(\phi)} g^E_{\mu\nu},\chi\Big). \label{BDActionE2}
\end{equation}
One can alternatively define Brans-Dicke theory starting from this last action,  by choosing the function $F(\phi(\phi^E))=\exp(\mathfrak g \phi^E)$. According to \cref{JtoEscalarBD} this corresponds to identifying
\begin{equation}
\mathfrak g^2=\frac {2}{3+2\omega}.
\end{equation}
The action \eqref{BDActionE2} leads to the simple equations of motions
\begin{align}
&\Box \phi^E \pm \frac{\mathfrak{g}}{M^2}T^E=0,\\
&R^E_{\mu\nu}-\p_{\mu}\phi^E\p_{\nu}\phi^E=\frac{1}{M^2}\Big(T^E_{\mu\nu}-\frac 12 g^E_{\mu\nu}T^E\Big),
\end{align}
where $T^E_{\mu\nu}=\frac{1}{F}\,T_{\mu\nu}$ is the Einstein frame stress-energy tensor. These are equivalent to \cref{scalar-eomBDJ,metric-eomBDJ} with constant $w$.

\paragraph{Axio-dilaton theory} We follow \cite{Burgess:2021qti} and consider the axion-dilaton theory in the Einstein frame as a case with K\"ahler target-space manifold. It can thus be described in terms of a complex scalar field $t=\frac 12(\tau+ia)$ and its complex conjugate $\bar{t}=\frac 12(\tau-ia)$, with only non-vanishing target-space metric and connection components reading 
\begin{equation}
\Gc^E_{t\bar t}=\frac 3{(t+\bar t)^2}=\frac 3{\tau^2}, \qquad \Gamma^{E\,t}_{\,\,\,\,\,tt}=-\frac{2}{t+\bar t}=-\frac 2\tau, \qquad \Gamma^{E\,\bar t}_{\,\,\,\,\,\bar t\bar t}=\Gamma^{E\,t}_{\,\,\,\,\,tt}.\label{functions_ADinEF}
\end{equation}
The scalars are coupled to gravity and matter through the defining functions
\begin{align}
F(t,\bar t)=t+\bar t=\tau, \qquad \Cc^t=\Cc^{\bar t}=0,
\end{align}
so that we simply get
\begin{equation}
  \qquad \p_{\bar t} F=1, \,\,\, \p^t F=\Gc^{E\, t b}\p_b F=\frac {(t+\bar t)^2}3=\frac {\tau^2}3, \qquad \mathcal{C}^{E\,t}=-\frac{F^t}{F}T^E=-\frac {\tau}3T^E. \label{FC_ADinEF}
\end{equation}
The Einstein frame action \eqref{ActionEF} thus reads:
\begin{align}
S&=\frac{M^2}2\int d^4x \sqrt{-g^E}\left [ R^E - \frac{6}{\tau^2}\partial_{\mu} t \, \partial_{\nu} \bar{t} \, g^{E\,\mu\nu} \right ]+ \int d^4x \, \mathcal{L}_m\Big(\frac{g^E_{\mu\nu}}{\tau},\chi, \varphi^a\Big) \nonumber \\
&=\frac{M^2}2\int d^4x \sqrt{-g^E}\left [ R^E - \frac{3}{2\tau^2} \left(\partial_{\mu} \tau \, \partial_{\nu} \tau + \partial_{\mu} a \, \partial_{\nu} a \right)  \, g^{E\,\mu\nu} \right ]+ \int d^4x \, \mathcal{L}_m\Big(\frac{g^E_{\mu\nu}}{\tau},\chi, \varphi^a\Big). \label{ActionAxioDilatonEF}
\end{align}
The scalar field equations \eqref{scalar-eom2EF}  simply read
\begin{equation}
\Box^E t - \frac{2}{t+\bar t} \p t\p t - \frac{t+\bar t}{6M^2}T^E=0,
\end{equation}
together with its complex conjugate, which contains the same informations. Their real and imaginary parts lead to
\begin{align}
&\Box^E \tau- \frac{1}{\tau} \Big[ \p \tau \p \tau-\p a \p a\Big] - \frac{\tau}{3M^2}T^E=0,\label{dilatonEFeom}\\
&\Box^E a - \frac{2}{\tau} \p \tau \p a=0. \label{axionEFeom}
\end{align}
The theory can also be expressed ignoring the K\"ahler structure by considering directly the scalar fields $\tau$ and $a$, with target-space functions
\begin{align}
& F(\tau,a)=\tau, \quad \mathcal{G}^E_{\tau\tau}=\mathcal{G}^E_{aa}=\frac{3}{2\tau^2}, \quad \mathcal{C}_{\tau}=\mathcal{C}_a=0, \nonumber\\
& \Gamma^{E\,a}_{\,\,\,\,\, \tau a}=-\frac{1}{\tau}, \quad \Gamma^{E\, \tau}_{\,\,\,\,\,a a}=-\Gamma^{E\,\tau}_{\,\,\,\,\, \tau \tau}=\frac{1}{\tau}, \quad F_{\tau}=1, \quad F_a=0, \quad F^{\tau}=\frac{2 \tau^2}{3}, \quad \mathcal{C}^{E \, \tau} = -\frac{2\tau}{3} T^E. \label{TSfunctionsEFAxioDReal}
\end{align}
Applying \cref{scalar-eom2EF} with the above functions expressed in terms of the real scalars directly gives back the equations of motion \eqref{dilatonEFeom} and \eqref{axionEFeom}.

\section{Parametrized post-Newtonian formalism and parameters}\label{section:PPNsection}

The parametrized post-Newtonian (PPN) formalism was developed to test predictions of metric theories of gravity in the weak-field and slowly-varying regime, hence directly applicable to solar-system experiments. This formalism allows one to compare metrics generated by matter sources in different theories of gravity. To do so, one computes the metrics generated by a perfect fluid matter source, by solving the field equations of the theory at post-Newtonian order. Once the metric is expanded in the parametrized post-Newtonian form, the comparison from one theory of gravity to another is done by looking at the expansion coefficients, called PPN parameters.

Metric theories of gravity postulate that matter and non-gravitational fields interact with the space-time metric only, forbidding any direct couplings with the other gravitational fields of the theory (which nevertheless play a role in the production of the metric). The goal of the present work is specifically to study the consequences of a (small) violation of this assumption, by introducing direct couplings between matter and gravitating scalar fields. One might wonder if this rules out the use of the PPN formalism to study the quasi-stationary weak-field regime. In the case where these direct couplings are strong, one certainly expects that matter dynamics would be ruled by interactions with the additional gravitating fields, rather than by the metric and hence the space-time geometry. Nevertheless, if these couplings are small enough, this shall not be the case, and matter could only be sensitive to the metric. In this latter case, it is thus of great interest to see how the presence of direct couplings between matter and additional gravitational fields affect the metric generated by localized matter sources. The PPN formalism is thus the ideal framework to study this question.

The goal of this section is to compute the PPN parameters in the presence of direct couplings between gravitating scalars and matter fields and investigate how they differ from the ones obtained without direct couplings. We will use the formalism and conventions exposed in \cite{Will:1993ns}.

\subsection{Parametrized post-Newtonian expansion of fields}

The PPN formalism studies an isolated post-Newtonian system in a homogeneous isotropic expanding universe. Matter sources of the system are modeled as perfect fluids. Each fluid element of speed $\vec v$ is made of matter with rest-mass density $\rho$, under pressure $p$. To apply the formalism, we will first perform the post-Newtonian expansion of the various fields of the theory. This amounts to expand the different gravitating fields (metric, scalar fields) generated by the matter source, in terms of the parameter:
 \begin{equation}
 \epsilon \equiv\frac{v}{c}. \label{epsilon}
\end{equation}
 Here $v^2=v^iv_i$ is the velocity of the fluid in the local quasi-Cartesian coordinates, also called standard post-Newtonian coordinates, which are asymptotically Minkowski. This coordinate system is chosen so that regions far from the isolated PN system are free falling in the cosmological model and at rest with respect to the Universe rest frame. One should thus expect a FLRW asymptotic form for the metric. However, as the time scale of the expansion of the Universe is today way smaller than the solar-system times scales, one can always use coordinates which are asymptotically Minkowski during the period of the experiment. We will thus generically denote the metric as: 
 \begin{equation}
g_{\mu\nu}\equiv\eta_{\mu\nu}+h_{\mu\nu}=\eta_{\mu\nu}+O(\epsilon^2), \label{hmunu}
\end{equation}
  where $\eta_{\mu\nu}$ is the Minkowski metric.
 
 \paragraph{PPN parameters and coefficients} The idea of the PPN formalism is to expand the fields of the theory in terms of functionals of the matter source, hence constructed from $\rho(x^{\mu})$,  $p(x^{\mu})$ or $\vec{v}(x^{\mu})$, $|\vec{x}-\vec{x'}|$\ldots There is {\it a priori} an infinity of possible functionals even at first PN order, hence at the $\epsilon^2$ or $\epsilon^4$ order.  Nevertheless, the PPN formalism restricts the choice of functionals. They should indeed satisfy various obvious conditions: they should vanish far from the source, they should have appropriate Lorentz transformations, they should not have reference to any preferred spatio-temporal origins. The PPN formalism moreover imposes slightly more arbitrary conditions: these functionals should not involve gradients of the matter sources functions and they should be rather simple \cite{Will:1993ns}. 
 
 The appropriate functionals are called PPN potentials or PPN functionals. Definitions and relations between some of these functionals are given in \cref{appendix:PNPotentials}. The metric should thus be expanded at first PN order in terms of the Newtonian potential $U$, post-Newtonian potentials $\Phi_W$, $\Phi_1$, $\Phi_2$, $\Phi_3$, $\Phi_4$, $\mathcal{A}$, $\mathcal{B}$ and functionals $V_i$,  $W_i$.  Once put in the so-called post-Newtonian gauge, where the spatial components of the metric are diagonal and the temporal component does not contain the $\mathcal{B}$ functional\footnote{For theories without time diffeomorphism invariance, such as Ho\v{r}ava gravity (in the unitary gauge), the functional $\mathcal{B}$ should also be kept in the PPN metric~\cite{Lin:2013tua}.}, it reads:
\begin{align}
&g_{00}=-1+2GU-2\beta U^2-2\xi \Phi_W+(2\gamma+2+\alpha_3+\zeta_1-2\xi)\Phi_1 \label{metricPPNexpansion_00} \nonumber\\
&\hspace{28pt}+2(3\gamma-2\beta+1+\zeta_2+\xi)\Phi_2+2(1+\zeta_3)\Phi_3+2(3\gamma+3\zeta_4-2\xi)\Phi_4-(\zeta_1-2\xi)\mathcal{A},\\
&g_{0j}=-\frac{1}{2}(4\gamma+3+\alpha_1-\alpha_2+\zeta_1-2\xi)V_j-\frac{1}{2}(1+\alpha_2-\zeta_1+2\xi)W_j, \label{metricPPNexpansion_0i}\\
&g_{jk}=(1+2\gamma U)\delta_{jk}.\label{metricPPNexpansion_jk}
\end{align}
We see that the component of the post-Newtonian metric depends only on the value of the Newton constant $G$ and ten parameters $\gamma$, $\beta$, $\xi$, $\alpha_1$, $\alpha_2$, $\alpha_3$, $\zeta_1$, $\zeta_2$, $\zeta_3$, $\zeta_4$. This is why this formalism is called parametrized post-Newtonian (PPN) formalism and the parameters are called PPN parameters. Of course, the expansion of \cref{metricPPNexpansion_00,metricPPNexpansion_0i,metricPPNexpansion_jk} directly gives the corresponding one for $h_{\mu\nu}$ defined in \cref{hmunu}.

The Newton constant $G$ is related, in a given theory, to a mass scale of the theory, e.g. $M$ in \cref{Action}, through the field equations. One often chooses the units with $M=1$ or the ones with $G=1$. This last condition can be imposed today through the cosmological matching conditions, explained below. Indeed, it will in general depend on the values of the fields far from the PN source, as can be seen from \cref{GPPN} below.

The $\gamma$ and $\beta$ parameters are already included in the Eddington-Robertson-Schiff (ERS) formalism \cite{1923mtr..book.....E,1962saa..conf..228R,1967rta1.book..105S}, while the other ones were defined in the full PPN formalism developed in \cite{Will:1971zzb,Will:1972zz,Will:1973zz}. The parameter $\gamma$ evaluates the quantity of space-time curvature produced by rest masses while  $\beta$ accounts for non-linearities present in the superposition law for gravity. In general relativity, both take unit values 
\begin{equation}
\gamma_{GR}=\beta_{GR}=1.
\end{equation}
The other PPN parameters also have a meaning in the standard PPN gauge. The parameter $\xi$ evaluates preferred-location effects, $\alpha_i$ evaluates preferred-frame effects, while $\alpha_3$ and $\zeta_i$ indicate violations of conservation of total momentum. For a detailed discussion, see \cite{Will:1993ns} and the summary of Table 4.3 therein. In our framework, we expect in advance non-vanishing values for the PPN parameters indicating violations of momentum conservation. Indeed, as will be explained around \cref{DmuTmunu}, the presence of direct couplings generically introduces terms in the divergence of the stress-energy tensor.

As for the metric, one shall also expand the gravitating scalar fields in terms of the $\epsilon$ order parameter:
\begin{equation}
\vp^a=\vp^a_0+\vp^a_2+\vp^a_4+O(\epsilon^6), \quad\qquad \vp^a_0=cst, \quad \vp^a_2=O(\epsilon^2), \quad \vp^a_4=O(\epsilon^4). \label{scalarPPN}
\end{equation}
The $\vp^a_0$, $\vp^a_2$ and $\vp^a_4$ are respectively of order $\epsilon^0$, $\epsilon^2$ and $\epsilon^4$. The latter are thus expanded on the PN potentials as
\begin{align}
&\vp^a_2=2\gamma_{\vp^a}U, \label{scalarPPNexpansion}\\
 &\vp^a_4=C^a_{UU} U^2+C^a_W \Phi_W + C^a_1 \Phi_1+C^a_2 \Phi_2+C^a_3 \Phi_3+C^a_4 \Phi_4+C^a_{\mathcal{A}} \mathcal{A}+C^a_{\mathcal{B}}\mathcal{B}. \label{scalarPPNexpansion2}
\end{align}

\paragraph{Cosmological matching conditions}The order $\epsilon^0$ fields are determined by the cosmological boundary conditions. Indeed, as the PN functionals vanish when the distance $r$ from the source goes to large values, the asymptotic values of the scalars are given by $\vp^a(r\to \infty)=\vp^a_0$. The constants $\vp^a_0$ shall thus be determined by the surrounding cosmological model, independent of the PN system, and will constitute cosmological matching conditions. We also recall that, as explained below \cref{epsilon}, the coordinate system is also chosen such that the metric approaches a Minkowski metric far from the source. In an expanding Universe this simply assumes that the typical time of gravitational experiments is small compared to cosmological times. 

\paragraph{Matter PN expansion} Finally, the matter sector should also be PN expanded. As already stated, one considers the perfect fluid approximation for the matter source, the stress-energy tensor of which thus reads:
\begin{equation}
T^{\mu\nu}=(\rho+\rho\Pi+p)u^{\mu}u^{\nu}+pg^{\mu\nu},
\end{equation}
where again, $\rho$ and $p$ are is the rest-mass energy density and pressure, $\rho\Pi$ the internal energy density, and $u^{\mu}=dx^{\mu}/d\tau$ the 4-velocity of the fluid elements normalised so that $u^{\mu}u_{\mu}=-1$. In solar systems, velocities of gravitating bodies are related to the Newtonian potential $U$ by virial relations. What's more,  their pressures $p$ and internal energies $\rho \Pi$ are smaller than their potential energy $\rho U$. The PN orders of the various functions defining the source are thus:
\begin{align}
&U\sim \rho\sim v^2 = O(\epsilon^2), \qquad p\lesssim \rho U = O(\epsilon^4), \qquad \Pi\lesssim U = O(\epsilon^2), \nonumber\\
& u^i=dt/d\tau dx^i/dt=u^0v^i=O(\epsilon).
\end{align} 
In the following perturbative study, we will use the fact that, in the quasi-static regime, temporal derivatives are smaller than gradients. This amounts to consider the relation
\begin{equation}
\frac{\partial}{\partial t}\sim \vec{v}\cdot \vec{\nabla} \sim O(\epsilon) \times \frac{\partial}{\partial x_i},
\end{equation}
when evaluating the PN order of a expression. As can be seen from \cref{metricPPNexpansion_00,metricPPNexpansion_0i,metricPPNexpansion_jk}, at lowest orders in $\epsilon$ the metric takes the form:
\begin{equation}
g_{00}=-1+2G U + O(\epsilon^4), \qquad g_{0j}=0 + O(\epsilon^4) ,\qquad g_{ij}=\delta_{ij}(1+2\gamma U) + O(\epsilon^4).
\end{equation}
Using the normalisation condition for $u^{\mu}$, the stress-energy tensor can thus be expanded to $O(\epsilon^4)$ as
\begin{equation}
T^{00}=\rho+\rho\Pi + \rho v^2 + 2 G \rho U + O(\epsilon^6), \qquad T^{ij}=\rho v^iv^j + p\delta^{ij} +O(\epsilon^6). \label{Tmunuexpansion}
\end{equation}
The coupling functions $\Cc^a$ must also be specified, for instance through their computation from a concrete matter Lagrangian $\mathcal{L}_m$, as well as PN expanded. Following the perfect fluid form for the stress-energy tensor, we shall simply expand them in the following way:
\begin{equation} 
\Cc^a= c^a \rho + c_{\Pi}^a \, \rho\Pi+ c_{v}^a \, \rho v^2  +  c_{U}^a \, \rho U  + c_{p}^a\,  p + O(\epsilon^6) \label{ccaexpansion}.
\end{equation}

In general the coupling functions $\Cc^a$ and their expansion \eqref{ccaexpansion} can be different for the source and the probe of gravity. 
As shown in subsection~\ref{subsec:redef-Jordanframe}, one can eliminate one among the three source coupling coefficients $c^a$, $c_{\Pi}^a$, $c_{p}^a$, by redefinition of the Jordan frame. Generally one cannot eliminate more than one of these coefficients.  At the moment, we however keep all of them until we derive the formulae for the PPN parameters. 

\subsection{PPN expanded field equations} 

Once every field has been PN expanded as in the previous section, the PPN expansion coefficients and the PPN parameters are to be computed through the field equations of the theory. They will be related to the target space functions $F$, $\Gc_{ab}$ and their derivatives. In order to solve perturbatively the field equations \eqref{scalar-eom2} and \eqref{metric-eom2}, we must thus PN expand them keeping track of the $\epsilon$ order of each term. To do so, we thus expand the target-space functions, giving for instance:
\begin{align}
F(\vp^a)=F(\vp^a_0)+F_b(\vp^a_0) \vp^b_2 + O(\epsilon^4), \qquad F_a(\vp^c)=F_a(\vp^c_0)+F_{ab}(\vp^c_0) \vp^b_2 + O(\epsilon^4).
\end{align}
as well as operators depending on the metric, such as the D'Alembertian operator $\Box$ or the (space-time) Levi-Civita connection:
\begin{align}
&\Box\vp^a|_2=g_0^{\mu\nu}\p_{\mu}\p_{\nu}\vp^a_2, \nonumber\\
&\Box \vp^a|_4=(g^{\mu\nu}\nabla_{\mu}\p_{\nu} \vp^a)|_4=g_2^{\mu\nu}(\nabla_{\mu}\p_{\nu} \vp^a)|_2 \, + \, g_0^{\mu\nu}(\nabla_{\mu}\p_{\nu} \vp^a)|_4 \nonumber \\
&\hspace{3.8cm}=g_2^{\mu\nu}\p_{\mu}\p_{\nu}\vp^a_2+g_0^{\mu\nu}\p_{\mu}\p_{\nu}\vp^a_4-g_0^{\mu\nu}(\Gamma^{\alpha}_{\mu\nu})_2\p_{\alpha} \vp^a_2, \label{BoxPPN4}\\
& (\Gamma^{\alpha}_{\mu\nu})_2=\frac 12 g_0^{\alpha\sigma}(\p_{\mu}g_{2\sigma\nu}+\p_{\nu}g_{2\sigma\mu}-\p_{\sigma}g_{2\mu\nu}).
\end{align}
The scalar equations \eqref{scalar-eom2} are thus expanded as
\begin{align}
&O(\epsilon^2): \, \Box\vp^a|_2 + \frac 32 A^a_b \Box\vp^b|_2=\frac{1}{2M^2}\left(B^a T_2- {\Cc^a_2}\right) ,  \label{SEPPN2}\\
&O(\epsilon^4): \, \Box\vp^a|_4 + \frac 32 A^a_b \Box\vp^b|_4 + C^a_{bc} \p_{\mu}\vp^b_2 \p_{\nu}\vp^c_2g_0^{\mu\nu}+\p_cA^a_{b} \vp^c_2 \Box\vp^b|_2 =\frac{\p_bB^a}{2M^2} \vp^b_2 T_2+\frac{1}{2M^2} (B^aT_4- {\Cc^a_4}), \label{SEPPN4}
\end{align}
where we defined
\begin{equation}
 A^a_b\equiv\frac{F^aF_b}{F}=B^aF_b, \quad B^a\equiv\frac{F^a}{F}, \qquad C^a_{bc}=\Gamma^a_{bc}+\frac{B^a}2(\Gc_{bc}+3F_{bc}). \\
\end{equation}
When not specified, as in \cref{SEPPN2,SEPPN4}, the above target-space functions and their derivatives are evaluated at $\vp^a_0$, being understood that the PN expansion of their variables has already been performed.

The trace-reversed Einstein equation \eqref{metric-eom2} is expanded as
\begin{align}
&O(\epsilon^2): \, FR_{2\mu\nu}-F_a\Big(\p_{\mu}\p_{\nu}\vp^a_2+\frac12 g_{0\mu\nu}\Box\vp^a|_2\Big)=\frac1{M^2}\Big(T_{2\mu\nu}-\frac 12 g_{0\mu\nu}T_2\Big),\\
&O(\epsilon^4):\, FR_{4\mu\nu}+F_a\vp^a_2R_{2\mu\nu}-F_{ab}\vp^b_2\Big(\p_{\mu}\p_{\nu}\vp^a_2-\frac12 g_{0\mu\nu}\Box\vp^a|_2\Big)\nonumber\\
&\hphantom{O(\epsilon^4):\,\,\, } -F_a\Big(\nabla_{\mu}\p_{\nu}\vp^a|_4+\frac12 g_{2\mu\nu}\Box\vp^a|_2+\frac12 g_{0\mu\nu}\Box\vp^a|_4\Big)\nonumber \\
&\hphantom{O(\epsilon^4):\,\,\, }  -\frac12 g_{0\mu\nu}F_{ab}\,  \p_{\alpha}\vp^a_2 \, \p_{\beta}\vp^b_2\, g_0^{\alpha\beta} - \Big(\Gc_{ab}+F_{ab}\Big)\p_{\mu}\vp^a_2\p_{\nu}\vp^b_2=\frac1{M^2}\Big(T_{4\mu\nu}-\frac 12 g_{2\mu\nu}T_2-\frac 12 g_{0\mu\nu}T_4\Big). 
\end{align}

\subsection{Determining the PPN parameters: procedure and results}\label{PPNcomputation}

The metric PPN parameters and the PPN coefficients of the various gravitational fields can be obtained by solving the field equations in a systematic way. We follow a procedure inspired by \cite{Will:1993ns}, that we summarize hereafter. 

The initial steps of the standard procedure of \cite{Will:1993ns}, corresponding to the determination of the gravitating and matter variables, the definition of the cosmological matching conditions and the PN expansion of the fields and equations of motion, have been explicitly showed earlier. Our procedure thus follows the successive steps:
\begin{itemize}
\item[] \underline{\it Step 1:} Solve for $h_{00}$ and $\varphi^a_2$ at order $O(\epsilon^2)$ using the $O(\epsilon^2)$ field equations. According to the expansions given in \cref{metricPPNexpansion_00,scalarPPN,scalarPPNexpansion}, this will thus determine the value of $G$ and $\gamma_{\varphi^a}$ in terms of the cosmological matching parameters. 
\item[] \underline{\it Step 2:} Solve for $h_{ij}$ at $O(\epsilon^2)$.  As can be seen from  \cref{metricPPNexpansion_jk}, this immediately gives the value of the $\gamma$ parameter. 
\item[] \underline{\it Step 3:} Solve for $h_{0i}$ at $O(\epsilon^3)$. From  \cref{metricPPNexpansion_0i}, we see that, comparing the $V_i$ and $W_i$ parameters in $h_{0i}$, one obtains the value of $\alpha_1$.
\item[] \underline{\it Step 4:} Solve for $\varphi^a_4$ using the scalar field equations \cref{SEPPN4}. According to \cref{scalarPPN,scalarPPNexpansion2}, this will give the $C^a_{UU}, C^a_{W}, C_i^a, C^a_{\mathcal A}, C^a_{\mathcal B}$ coefficients in terms of the scalar target-space functions and the expansion coefficients of the coupling $\mathcal{C}^a$ shown in \cref{ccaexpansion}.
\item[] \underline{\it Step 4:} Solve for $h_{00}$ at $O(\epsilon^4)$ using the field equations at order $O(\epsilon^4)$. This is the most computationally expansive step. It leads to the obtention of the remaining PPN coefficients, namely $\alpha_2$, $\alpha_3$, $\alpha_4$, $\beta$, $\xi$, $\zeta_1$, $\zeta_2$, $\zeta_3$.
\end{itemize} 

In order to implement the above procedure, we transform the differential equations for the fields into algebraic equations for the PPN coefficients. To this end, we first plug the PPN expansions for the various fields into the field equations, then use the knowledge of specific relations between the derivatives of the PN potentials. Some of the latter are given in \cref{appendix:PNPotentials}. In this manner, solving for the various fields at a certain PN order $O(\epsilon^n)$ turns into solving the algebraic equations in terms of the PPN coefficients playing a role at this order.
We obtained, step by step, the following PPN parameters in terms of the target space functions:
\begin{align}
&\text{\underline{\it Step 1:}} \qquad &G&=\frac{(4B^c +{c^c} )F_c+2}{F (3B^cF_c+2)},  \label{GPPN} &&&\\
& &\gamma_{\varphi^a}&= \frac{F}{2}\times \frac{2(B^a+{c^a})-3(B^a{c^c}-B^c{c^a})F_c}{(4B^c +{c^c} )F_c+2}\nonumber&&&\\
& &&=\frac{1}{2G} \frac{2(B^a+{c^a})-3(B^a{c^c}-B^c{c^a})F_c}{3B^cF_c+2}. \label{gammaphiPPN} &&&\\
&\text{\it \underline{Step 2:}} \qquad &\gamma&=\frac{(2B^c-{c^c})F_c +2}{(4B^c+{c^c})F_c+2}=1-\frac{2(B^c +{c^c})F_c }{(4B^c +{c^c} )F_c+2}. \label{gammaPPN} &&&\\
&\text{\it \underline{Step 3:}} &\alpha_1&=0.&&\label{alpha1PPN}\\
&\text{\it \underline{Step 4 \& 5:}} &\alpha_2& =0, \qquad \alpha_3=0, \qquad  \xi=0, \qquad \zeta_1=0, \label{vanishingPPN}&&&\\
&&\zeta_2&=\frac{24 (B^a{c^c}-B^c{c^a})F_cF^b\tilde F_{ab}}{(3 B^cF_c+2)(4 B^cF_c+{c^cF_c}+2)^2} \nonumber &&& \\
&&&+\frac{  \Big(2(B^a+{c^a})-3(B^a{c^c}-B^c{c^a})F_c\Big) (F^bF_b{c_a}+FF^b\nabla_a{c_b}-F{c^b}\tilde{F}_{ab})}{(4 B^cF_c+{c^cF_c}+2)^2},\hspace{-1cm} \label{zeta2PPN}&&&\\
&&\zeta_3&=\frac{{(c^c_{\Pi}-c^c)} F_c}{4B^cF_c+{c^cF_c}+2}, \qquad \zeta_4=\frac{{(\frac 13 c^c_{p}+c^c)} F_c}{4B^cF_c+{c^cF_c}+2}, \label{zetaPPN}&&&\\
&&\beta&= 1+\frac{(B^c{+c^c)}(B^d{+c^d)}(F_cF_d-2F\tilde F_{cd})}{(3 B^cF_c +2)(4B^cF_c+{c^cF_c}+2)^2}, \label{betaPPN}&&&
\end{align}
where in the expression for $\beta$ and $\zeta_2$, we introduced the target space covariant tensor 
\begin{equation}
  \tilde F_{cd}\equiv\nabla_cF_d= \nabla_c\partial_dF=F_{cd}-\Gamma^e_{cd}F_e. 
\end{equation}
In the above expressions, in which we set $M\equiv1$, all the functions $F_c$, $B^c$, $F$, $c^c$ are evaluated at the cosmological background values  $\varphi^a_0$ for the gravitating scalars. As mentioned in the text below \cref{metricPPNexpansion_jk} we  thus see from \cref{GPPN} that the value for the Newton constant $G$ depends on the values of the cosmological fields $\varphi^a_0$.

The above expressions are thus the general expressions for the ten PPN parameters in multi-scalar theories of gravity with direct coupling. The can be used for any theory once put in the form of \cref{Action}. For multi-scalar theories of gravity without direct coupling, the $\beta$ and $\gamma$ parameters were already computed in \cite{Damour:1992we}, although with different notations. 

 One can see from the above formulae that in the special case where $B_c=-c_c$, the PPN parameters are identical to those of general relativity. This is not a surprise, as going back to the $O(\epsilon^2)$ order expansion \cref{SEPPN2} of the scalars field equations \cref{scalar-eom2}, we see that this corresponds to the case where the scalars decouple. This can also be seen easily from the computation \eqref{CcE} of the direct couplings in the Einstein frame.

In other cases, the PPN parameters are different. For small direct couplings, $i.e.$ for $c^a\ll B^a$, they differ only very slightly from the PPN parameters obtained in the corresponding theory without direct couplings. This seems to go against the claim made in \cite{Burgess:2021qti} for the axio-dilaton theory. We come back to this specific case below.

\subsection{Redefining the Jordan frame} \label{subsec:redef-Jordanframe}
We remark that, at the first PPN order, most of the PPN parameters obtained in \cref{GPPN,gammaphiPPN,gammaPPN,alpha1PPN,vanishingPPN,zeta2PPN,zetaPPN,betaPPN} depend on the direct couplings only through $c^a$. We will show that a Weyl transformation allows to go in a frame where this coefficient vanishes. Using \cref{Tmunuexpansion} we first expand the trace of the stress-energy tensor:
\begin{equation}
T=T^{\mu\nu}g_{\mu\nu}=-\rho -\rho\Pi + 3p + O(\epsilon^6),
\end{equation}
and use its expression to write the coupling $\mathcal{C}^a$ of \cref{ccaexpansion} as
\begin{equation}
\mathcal{C}^a=-c^a T + (c^a_{\Pi}-c^a) \rho \Pi + c^a_v \rho v^2 + c^a_U \rho U +( c^a_p + 3 c^a)p + O(\epsilon^6). \label{CafromT}
\end{equation}
When can then apply a Weyl transformation on the initial action \eqref{Action}, similar to the one used in \cref{JFtoEF} to go to Einstein frame, but here with generic function $f(\varphi^a)$. Defining the new metric $\bar{g}$ through 
\begin{equation}
g_{\mu\nu}=\frac{1}{f(\varphi^a)}\bar{g}_{\mu\nu},
\end{equation} 
the new target-space functions, stress-energy tensor and direct couplings read
\begin{align}
&\bar{F}=f F, \qquad \bar{\mathcal{G}}_{ab}=f \mathcal{G}_{ab}+\frac{3}{2} F f_a f_b + 3 F_a f_b, \qquad f_a\equiv \frac{\p f}{\p \varphi^a}, \,\, F_a\equiv \frac{\p F}{\p \varphi^a}, \label{Newframefunctions}\\
&\bar{T}^{\mu\nu}=f^3T^{\mu\nu}, \qquad \bar{T}=f^2 T, \qquad \bar{\mathcal{C}}^a=f^2 \mathcal{C}^a + f f_a T. 
\end{align}
From this last equality and from \cref{CafromT}, we deduce that by choosing 
\begin{equation}
f(\varphi^a)=e^{c_a \varphi^a}, \label{Weyltranscouplings}
\end{equation}
we can go in a frame with
\begin{align}
&\bar{\mathcal{C}}^a=f^2 \left\{  (c^a_{\Pi}-c^a) \rho \Pi + c^a_v \rho v^2 + c^a_U \rho U +( c^a_p + 3 c^a)p \right\} + O(\epsilon^6).
\end{align}
In this frame, the new expansion parameters thus read
\begin{equation} 
\bar{c}^a=0, \quad \bar{c}^a_{\Pi}=f^2 (c^a_{\Pi}-c^a), \quad  \bar{c}^a_v=f^2 c^a_v, \quad \bar{c}^a_U=f^2 c^a_U, \quad \bar{c}^a_p=f^2( c^a_p + 3 c^a),
\end{equation}
As motivated above, we see that in this frame $\bar{c}^a=0$ so that the PPN parameters of \cref{GPPN,gammaphiPPN,gammaPPN,alpha1PPN,vanishingPPN,zeta2PPN,zetaPPN,betaPPN} take the simpler form
\begin{align}
& &G&=\frac{4\bar B^c \bar F_c+2}{\bar F (3\bar B^c\bar F_c+2)},  \label{GPPNbar} &&&\\
& &\gamma_{\varphi^a}&=\frac{\bar F \bar B^a}{4\bar B^c\bar F_c+2}=\frac{1}{G} \frac{\bar B^a}{3\bar B^c\bar F_c+2}. \label{gammaphiPPNbar} &&&\\
& &\gamma&=\frac{2\bar B^c\bar F_c +2}{4\bar B^c\bar F_c+2}=1-\frac{2\bar B^c\bar F_c }{4\bar B^c\bar F_c+2}. \label{gammaPPNbar} &&&\\
& &\alpha_1&=0, \qquad  \alpha_2=0, \qquad \alpha_3=0, \qquad  \xi=0, \qquad \zeta_1=0, \quad \zeta_2=0 \label{vanishingPPNbar}&&&\\
&&\zeta_3&=\frac{\bar c^c_{\Pi} F_c}{4\bar B^c\bar F_c+2}, \qquad \zeta_4=\frac 13 \frac{{\bar c^c_{p}} \bar F_c}{4\bar B^c\bar F_c+2}, \label{zetaPPNbar}&&&\\
&&\beta&= 1+\frac{\bar B^c\bar B^d(\bar F_c\bar F_d-2\bar F\tilde{\bar F}_{cd})}{(3 \bar B^c\bar F_c +2)(4\bar B^c\bar F_c+2)^2}. \label{betaPPNbar}&&&
\end{align}
We see that these PPN parameters differ from the ones of the multi-scalar tensor theory defined with $\bar F$, $\bar{\mathcal{G}}_{ab}$ only through  $\zeta_3$ and $\zeta_4$. Hence, the Weyl transformation of \cref{Weyltranscouplings} allows to go in a new Jordan frame, where the direct couplings between the gravitating scalars and the source are at least of $\epsilon^4$ order, and the defining functions of the gravitating part of the theory are given by \cref{Newframefunctions}. Depending on the case, it might be easier to extract the PPN parameters from the initial action \cref{Action} through the formulae given in  \cref{GPPN,gammaphiPPN,gammaPPN,alpha1PPN,vanishingPPN,zeta2PPN,zetaPPN,betaPPN}, or perform first the above Weyl transformation and then use expressions of \cref{GPPNbar,gammaphiPPNbar,gammaPPNbar,vanishingPPNbar,zetaPPNbar,betaPPNbar}.

\subsection{The PPN parameters for particular cases}

We now apply the above results for the PPN parameters to the two examples presented in \cref{subsection:particularcases}. The first example can be seen as a consistency check of our formulae. On the other hand, the second example is one of the motivations of this work. 
 
\paragraph{Brans-Dicke theory} The Brans-Dicke theories presented in \cref{subsection:particularcases} correspond to the target-space functions: 
\begin{align} 
&F(\vp^a)=\phi, \quad \Cc_a=0, \quad F_a=1, \quad F_{ab}=\p_{ab}F=0, \quad F^a=\Gc^{ab}{F}_b=\frac\phi \omega, \\
&\tilde F_{ab}=-\Gamma^c_{ab}F_c=\frac 12 \Big(\frac 1\phi-\frac{\omega'}\omega\Big).
\end{align}
We thus deduce the following expressions for the relevant terms
\begin{align}
&B^a=\frac{F^a}{F}=\frac{1}{\omega},\qquad B^cF_c= \frac{1}{\omega}, \qquad 2 B^cB^d F \tilde F_{cd}= \omega^{-2} \Big(1-\phi \frac{\omega'}\omega\Big)
\end{align}
appearing in expressions \eqref{GPPN} to \eqref{betaPPN} for the PPN parameters. The non-vanishing PPN parameters thus read
\begin{align}
&\gamma_{\phi}=\frac{F}{2}\times \frac{B^a}{2B^c F_c+1}=\frac{\phi_0}{2(\omega+2)},\, \qquad G=\frac{4B^cF_c+2}{F (3B^cF_c+2)}=\frac{4+2\omega}{\phi_0(3+2\omega)}, \\
 &\gamma=\frac{B^cF_c +1}{2B^cF_c+1}=\frac{1+\omega}{2+\omega}, \quad \beta-1=\frac{B^cB^d(F_cF_d-2F\tilde F_{cd})}{4(3 B^cF_c +2)(2B^cF_c+1)^2}=\frac{\phi_0 \, \omega'}{4(3+2\omega)(2+\omega)^2}.
 \end{align}
As should be clear from the discussion on the PPN formalism, in these results the values for the target-space functions are evaluated for the cosmological background. One should thus read $\omega=\omega(\phi_0)$ and $\omega'=\omega'(\phi_0)$. From the standard results above, we see that for a given function $\omega(\phi)$, namely a given Brans-Dicke theory, constraints on the PPN parameters directly fix the maximal possible strength of coupling between matter and the scalar field $\phi_0$.

\paragraph{Axio-dilaton theory} This case was defined in \cref{functions_ADinEF} in the Einstein frame following the study of \cite{Burgess:2021qti}. The PPN parameters being computed in the Jordan frame, one has to go from the target-space metric in the Einstein frame to the one in the Jordan frame in order to apply the above formulas. Inverting \cref{relationTSmetrics} the Jordan frame target-space metric is expressed through the Einstein frame one by:
\begin{equation}
\mathcal{G}_{ab}=F\mathcal{G}^E_{ab}-\frac{3}{2}\frac{F_aF_b}{F}.
\end{equation}
Note that when going to Jordan frame, the K\"ahler structure might not be preserved. Hence it is necessary to work with real scalars. We thus deduce the Jordan frame target-space functions of the axio-dilaton case from \cref{TSfunctionsEFAxioDReal}:
\begin{align}
& F(\tau, a)=\tau \quad F_{\tau}=1,\quad  F_a=0,  \label{Ffunctionaxiodilaton}\\
& \mathcal{G}_{\tau\tau}= \tau \mathcal{G}^E_{\tau\tau} -\frac{3}{2\tau}=0, \quad \mathcal{G}_{aa}=\tau \mathcal{G}^E_{aa}=\frac{3}{2\tau}. \label{MetriaxiodilatonDegen}
\end{align}
Here, the subscript $a$ refers to the axion field $a(x^{\mu})$, as shall be clear to the reader. From \cref{MetriaxiodilatonDegen} we see that this case is slightly degenerate because of the vanishing of the dilaton part of the target-space metric, $i.e$ the absence of dilaton kinetic term in the Jordan frame. The action \eqref{ActionAxioDilatonEF} is indeed written in the Jordan frame as
\begin{equation}
S=\frac{M^2}2\int d^4x \sqrt{-g}\Big [\tau R -\frac{3}{2\tau}\p^{\mu} a \p_{\mu} a\Big ] + \int d^4x \, \mathcal{L}_m(g_{\mu\nu},\chi), \label{JFaxiodilaton}
\end{equation}
the dilaton part of which corresponds to a Brans-Dicke scalar with vanishing function $\omega(\tau)=0$. 

The PPN parameters can be  obtained by taking formally the limit $B^{\tau} = \mathcal{G}^{\tau\tau}F_{\tau} \to \infty$ in \cref{GPPN} to \eqref{betaPPN}: 
\begin{equation}
\gamma=\frac{1}{2}, \quad \beta=1, \quad
\alpha_1=\alpha_2=\alpha_3=\xi=\zeta_1=\zeta_2=\zeta_3=\zeta_4=0. \label{PPNaxiodilaton}
\end{equation}
This result naturally agrees with the Brans-Dicke case with $\omega=0$.

The degeneracy of the Jordan frame action \eqref{JFaxiodilaton} comes from the fact that the Einstein frame kinetic terms for the dilaton (and axion) are exactly the ones generated by the change in Ricci scalar under the Weyl transformation. Note that vanishing kinetic terms in the Jordan frame are harmless, as opposite to vanishing kinetic terms in the Einstein frame, which reveal strongly coupled dynamics. In order to obtain non-degenerate Jordan target-space functions, one could nevertheless consider slightly different normalization of the Einstein frame kinetic terms, while keeping the same gravity coupling function $F(\tau,a)=\tau$. For instance, one might parametrize kinetic terms for the dilaton $\tau$ through a small parameter $\epsilon_\tau$, by adding to the action \eqref{JFaxiodilaton} a term of the form
\begin{equation}
S_{\epsilon_{\tau}}=\frac{M^2}2\int d^4x \sqrt{-g}\Big [\frac{\epsilon_{\tau} }{2\tau}\p^{\mu} \tau \p_{\mu} \tau \Big]. \label{epsilonAction}
\end{equation}
The target-space metric \eqref{MetriaxiodilatonDegen} would then contain terms proportional to $\epsilon_\tau$, in particular a non-vanishing $\mathcal{G}_{\tau\tau}$. By first applying the PPN formulae \eqref{GPPN} to \eqref{betaPPN}, then taking the limit $\epsilon_{\tau} \rightarrow 0$, one obtains back the results of \cref{PPNaxiodilaton}.

\paragraph{Axio-dilaton with direct coupling} The authors of \cite{Burgess:2021qti} introduced a direct coupling to matter for the axion field $a(x^{\mu})$. It is defined in the Einstein frame as
\begin{equation}
\mathcal{A}\equiv\frac{2}{\sqrt{-g^E}}\frac{\delta S_m}{\delta a}.
\end{equation}
The coupling $\mathcal{A}$ should agree with the Einstein frame coupling $\mathcal{C}^E_a$. The latter were obtained through \cref{CcE} from $\mathcal{C}_a$, the Jordan frame couplings defined in \cref{def-Ca} . According to \cref{Ffunctionaxiodilaton}, in the present case we have $F_a=0$ for the axion, so that $\mathcal{A}$ is here identical to the Einstein frame coupling
\begin{equation}
\mathcal{A}\equiv\frac{2}{\sqrt{-g^E}}\frac{\delta S_m}{\delta a}=\frac{\sqrt{-g}}{\sqrt{-g^E}} \mathcal{C}_a= \mathcal{C}^E_a,
\end{equation}
as should be. The last equality is obtained by applying \cref{CcE} with $\varphi^a=a$ being the axion field. In order to evaluate the expressions for PPN coefficients derived in \cref{PPNcomputation}, we have to identify the Jordan frame couplings $ \mathcal{C}_c$ and their PPN expansion. They would thus read
 \begin{equation}
 \mathcal{C}_a=\frac{\sqrt{-g^E}}{\sqrt{-g}}\mathcal{A}=\tau^2 \mathcal{A}, \qquad  \mathcal{C}_{\tau}=0 \qquad \qquad \mathcal{C}^a=\mathcal{G}^{aa} \mathcal{C}_a=\frac{2 \tau^3}{3} \mathcal{A}, \quad \mathcal{C}^{\tau}=0. \label{Caaxiodilaton}
 \end{equation}
Following the initial work \cite{Burgess:2021qti}, we consider the simplest coupling expansion for non-relativistic sources, taking the form
\begin{equation}
\mathcal{A}=\epsilon_{cpl.} \rho, \label{Aepsilonrho}
\end{equation}
for constant $\epsilon_{cpl.}$ coupling coefficient. The cumbersome subscript is used to avoid confusion with the PN expansion parameter $\epsilon$. According to \cref{ccaexpansion,Caaxiodilaton} the above coupling corresponds to
\begin{equation}
  c_a= \tau^2 \epsilon_{cpl.}, \quad c^a=\frac{2}{3} \, \tau^3 \epsilon_{cpl.}, \qquad c_{\tau}=0, \quad c^{\tau}=0. \label{couplingsDirect}
\end{equation}
All the other coefficients of the PN expansion \cref{ccaexpansion}, such as $c^a_{\Pi}$ or $c^a_{p},$ vanish in this case. In \cref{couplingsDirect}, we should obtain $c^{\tau}$ by raising the index with the inverse metric $\mathcal{G}^{\tau\tau}$. Whereas the latter is ill defined, as $\mathcal{G}_{\tau\tau}$ vanishes, it is still natural to take $c^{\tau}=0$. To be convinced, one could go to the non-degenerate case by adding the term $\epsilon_\tau$ term of \cref{epsilonAction} to the action. One would then obtain a non-vanishing target-space metric element $\mathcal{G}_{\tau\tau}$, finite $\mathcal{G}^{\tau\tau}$, and indeed get $c^{\tau}=0$.

Now that we identified the coupling functions and their expansion parameters, we can use the expression found in \cref{PPNcomputation} for the PPN parameters in the presence of direct couplings. From \cref{gammaPPN} it appears that the $\gamma$ parameter is unchanged compared to the case without the direct couplings, as $c^cF_c=c^aF_a+c^{\tau}F_{\tau}=0$. Moreover, the $\beta$ parameter is also unchanged, because the direct couplings $c^c$ have no effect when taking the limit $B^c F_c \rightarrow \infty$. We thus obtain the same PPN parameters as in the case without the direct axion coupling:
\begin{equation}
\gamma=\frac{1}{2}, \quad \beta=1, \quad
\alpha_1=\alpha_2=\alpha_3=\xi=\zeta_1=\zeta_2=\zeta_3=\zeta_4=0. \label{againPPNaxiodilaton}
\end{equation}

More generally, we remark that when the function $F(\{\varphi^a\})$ and direct couplings $\mathcal{C}^{b}$ depend on sets of scalar fields $\{\varphi^a\}$ and $\{\varphi^b\}$ containing no common element, the parameter $\gamma$ is unaffected by the direct couplings since  $c^cF_c=0$. This is $a\, priori$ not the case for the parameter $\beta$  since there is a term in $\tilde{F}_{ab}$ containing the connection term $\Gamma^e_{ab}F_e$ which can mix the indices.

\paragraph{Remark on the computation of \cite{Burgess:2021qti}} By computing the first terms of the metric expansion in  a spherically symmetric case, the authors of this previous work derived formulae for PPN parameters, valid in the limit $\epsilon_{cpl.}\rightarrow 0$ where the axion direct coupling \eqref{Aepsilonrho} is small. According to  \textrm{(3.46) and (3.47) of \cite{Burgess:2021qti} } they read
\begin{equation}
\gamma=\frac{3-\epsilon_{cpl.}\beta \tanh \delta}{3+\epsilon_{cpl.}\beta \tanh \delta}, \qquad \beta=1+\frac{\epsilon_{cpl.}^2\beta^2}{9(\cosh \delta +\frac 13 \epsilon_{cpl.} \beta \sinh \delta)^2} , \label{theirPPN}
\end{equation}
where $\delta$ and $\beta$ are related to boundary conditions of the solutions, namely the asymptotic values of the field. Indeed, in their formulae ((3.37), (3.38) and (3.40) of \cite{Burgess:2021qti}), the authors of \cite{Burgess:2021qti} give the following relations:
\begin{align}
&\gamma_{b.c.}=\frac{-2\epsilon_{cpl.} GM}{3}, \qquad \alpha \,  \gamma_{b.c.} \approx \frac{-2GM}{3}, \label{relation1} \\
&a_{\infty}=\alpha-\beta \tanh \delta, \qquad \tau_{\infty}=\frac{\beta}{\cosh \delta} . \label{relation2}
\end{align}
From \cref{relation1} we directly deduce that 
\begin{equation}
\alpha=\frac{1}{\epsilon_{cpl.}}.
\end{equation}
The values of the axion and dilaton profiles far from the source, $i.e.$ $a_{\infty}$ and $\tau_{\infty}$, have to be finite and are fixed by the cosmological matching conditions even in the $\epsilon_{cpl.} \rightarrow 0$ limit. One can thus rewrite the first equality in \eqref{relation2} as
\begin{align}
&\beta \tanh \delta=\frac{1}{\epsilon_{cpl.}} - a_{\infty},
\end{align}
so that the two PPN parameters of \cref{theirPPN} can be expanded 
\begin{align}
&\gamma=\frac{ 2 - \epsilon_{cpl.}\, a_{\infty}}{4 +\epsilon_{cpl.}  a_{\infty}}= \frac{1}{2} + O(\epsilon_{cpl.}), \\ 
&\beta - 1= \frac{\epsilon_{cpl.}^2\beta^2}{9\cosh^2\delta(1+\frac 13 \epsilon_{cpl.} \beta \tanh \delta)^2}= \frac{\epsilon_{cpl.}^2\beta^2}{\cosh^2\delta(4 - \epsilon_{cpl.}  a_{\infty})^2}=\frac{\epsilon_{cpl.}^2 \tau_{\infty}^2 }{(4-\epsilon_{cpl.}  a_{\infty})^2}=O(\epsilon_{cpl.}^2).
\end{align}
We see that in the limit $\epsilon_{cpl.} \rightarrow 0$ of validity of these expressions, they simply lead to the Brans-Dicke PPN parameters with $\omega(\tau)=0$, as we found in \cref{againPPNaxiodilaton}. In a more recent work \cite{Brax:2022vlf} the same authors study the possibility to evade this fact by considering non-linear coupling for the axion, hence relaxing the condition \cref{Aepsilonrho}. 

\section{Observational constraints on the PPN parameters}\label{section:Observations}

In this section, we tackle the question of classical constraints on the PPN parameters in the presence of direct couplings. In order to obtain the equations for classical tests of gravity, and thus constrain the PPN parameters, the full approach should study the post-Newtonian equations of motion for a system of massive bodies constituted of a PN source and a probe, such as the Sun and a planet. In general, this must be done by treating the massive bodies as gravitating clusters of massive particles and obtaining the equations of motion for their centers of mass. This study is not straightforward in the presence of direct couplings, as the definition of mass densities is not obvious in the PPN formalism. Indeed, neither the rest-mass density nor the mass-energy density are conserved. However, by averaging on internal dynamical timescales, which is justified by the fact that the time scales of the changes on the internal structures of the Sun and planets are way shorter than the typical orbital times, the final equations of motions can be obtained without such precise considerations \cite{Will:1993ns}. The obtention of the equation for the acceleration of massive bodies, depending on the ten PPN parameters, is nevertheless not immediate. In the present case, the intermediate steps using continuity equations and conservation law integrals should take into account the presence of direct couplings.

Another approach is to consider the PPN metric generated by the PN source and looking at the test probe simply as a massive point particle. In this approach, the massive source and massive probe are treated differently in general. For instance, while a point particle is structureless, some of the post-Newtonian gravitational effects included in the PPN metric can be generated from rotation or non-sphericity of the source. Furthermore, in general the source and the probe of gravity may have different properties so that the coupling functions $\Cc^a$ and their expansion \eqref{ccaexpansion} can be different for the source and the probe. In particular, although in subsection~\ref{subsec:redef-Jordanframe} we have set $c^a=0$ for the source by redefinition of the Jordan frame, $c^a$ for the probe in this frame may be non-zero in general. In \cref{ppgeneraldynamics} we evaluate the possible modifications from geodesic motion of a test point particles, due to the presence of direct couplings responsible for additional forces.

If one simply wishes to access the $\gamma$ and $\beta$ parameters, it can be sufficient to obtain the equation of motions of point particles in the presence of a background generated by a spherical static matter source, as in the formalism developed by Eddington, Robertson and Schiff \cite{1923mtr..book.....E,1962saa..conf..228R,1967rta1.book..105S}. In \cref{ppparticulardynamics} we show that one can access the $\gamma$ and $\beta$ parameters through the classical experiments on massive bodies, such as the measurement of the perihelion shift of Mercury, even in the presence of small direct couplings. In fact, these experiments can be seen as constraints on the strength of these couplings.

Finally, in \cref{photonsdynamics} we motivate that photon dynamics also allow to access the $\gamma$ parameter, even in the presence of direct couplings, through classical tests measuring the  time delay or deviation of light.

\subsection{Direct coupling in the point-particle Lagrangian}\label{ppgeneraldynamics}

\paragraph{Divergence of the stress-energy tensor in field theory} We shall see below that the forces exerted on point particles are related to the divergence of the stress-energy tensor. Towards this goal, as a warm up, let us briefly review a well-known identity involving the divergence of the stress-energy tensor in field theory. For this purpose, we consider a diffeomorphism-invariant action of the form ${S}_m[\chi(x), \varphi^a(x), g_{\mu\nu}(x)]$, where $\chi(x)$, $\varphi^a(x)$ and $g_{\mu\nu}(x)$ represent the matter fields, the gravitational scalars and the metric, respectively. Under a generic change of coordinates $x^{\mu}\rightarrow x^{\mu} + \xi^{\mu}$, they transform as follows:
\begin{equation}
g_{\mu\nu}\rightarrow g_{\mu\nu} - \nabla_{\mu}\xi_{\nu} - \nabla_{\nu}\xi_{\mu}, \quad \varphi^a\rightarrow\varphi^a - \xi^{\mu}\partial_{\mu}\varphi^a, \quad \chi \rightarrow \chi - \mathcal{L}_{\xi}\chi = \chi - \xi^{\mu}\nabla_{\mu}\chi + d^{\nu}_{\,\,\,\mu} \nabla_{\nu} \xi^{\mu}. \label{transformation-gmunu-varphia-chi}
\end{equation}
The first two transformations are imposed by the transformation properties of the metric and the scalars while the last one is kept generic. The vanishing (up to a total derivative) of the variation of the matter action under such a diffeomorphism leads to the usual identity involving the divergence of the stress-energy tensor. It reads:
\begin{equation}
\nabla_{\nu}T^{\nu}_{\ \mu} = \frac{1}{2}\mathcal{C}_a\partial_{\mu}\varphi^a + \frac{1}{2}\mathcal{E}_{\chi}\nabla_{\mu}\chi + \frac{1}{2}\nabla_{\nu}(d^{\nu}_{\,\,\,\mu}\mathcal{E}_{\chi}), \label{nonconservation1}
\end{equation}
where we used the definition given in \cref{def-Ca} and
\begin{equation}
\mathcal{E} _{\chi} \equiv \frac{2}{\sqrt{-g}}\frac{\delta S_m}{\delta\chi}.
\end{equation}
When $\mathcal{C}_a=0$, the right-hand side of \cref{nonconservation1} vanishes on-shell, $i.e.$ upon using the equation of motion $\mathcal{E}_{\chi}=0$, leading to the conservation of the stress-energy tensor. In our case, only the last two terms vanish once we incorporate the matter equations of motion $\mathcal{E}_{\chi}=0$. Hence, the stress-energy tensor equation \eqref{nonconservation1} reduces to:
\begin{equation}
\nabla_{\nu}T^{\nu}_{\ \mu} = \frac{1}{2}\mathcal{C}_a\partial_{\mu}\varphi^a. \label{DmuTmunu}
\end{equation}

\paragraph{Divergence of the point particle stress-energy tensor} Let us now derive an identity similar to \eqref{nonconservation1} for a point particle action of the form $S_P[x_P^{\mu}(\lambda), \varphi^a(x), g_{\mu\nu}(x)]$, where $x_P^{\mu}(\lambda)$ denotes the worldline of the particle parametrized by the parameter $\lambda$, $\varphi^a(x)$ the gravitational scalars and $g_{\mu\nu}(x)$ the metric. Under the infinitesimal diffeomorphism transformation $x^{\mu}\rightarrow x^{\mu} + \xi^{\mu}(x)$ we have, in addition to the first two transformations rules of (\ref{transformation-gmunu-varphia-chi}), the following one for the  $x_P^{\mu}(\lambda)$ worldline:
\begin{equation}
 x_P^{\mu}(\lambda) \to x_P^{\mu}(\lambda) + \xi^{\mu}(x_P(\lambda)).
\end{equation}
Hence, the action varies under an infinitesimal diffeomorphism as
\begin{equation}
\delta S_P  = \int d\lambda \frac{\delta S_P}{\delta x_P^{\mu}(\lambda)}\xi^{\mu}(x_P(\lambda)) + \int d^4x\sqrt{-g}  \left[ -\frac{1}{2}\mathcal{C}_a\partial_{\mu}\varphi^a + \nabla_{\nu}T^{\mu\nu}\right] \xi^{\mu}.
\end{equation}
By requiring the diffeomorphism invariance of the action, $i.e.$ by setting this variation to zero for arbitrary $\xi^{\mu}(x)$ parameter, one obtains the following identity
\begin{equation}
 \nabla_{\nu}T^{\mu\nu} = \frac{1}{2}\mathcal{C}_a\partial_{\mu}\varphi^a
  - \int d\lambda \frac{\delta S_P}{\delta x_P^{\mu}(\lambda)} \frac{\delta^4(x-x_P(\lambda))}{\sqrt{-g}}.
\end{equation}
Upon using the equation of motion $\delta S_P/\delta x_P^{\mu}(\lambda)=0$ for $x_P^{\mu}(\lambda)$, we again obtain 
\begin{equation}
\nabla_{\nu}T^{\nu}_{\ \mu} = \frac{1}{2}\mathcal{C}_a\partial_{\mu}\varphi^a. \label{DmuTmunu-pp}
\end{equation}

We can find another identity from the invariance of the action $S_P[x_P^{\mu}(\lambda), \varphi^a(x), g_{\mu\nu}(x)]$ under reparametrizations of the worldline parameter $\lambda\to \lambda + \zeta(\lambda)$. The functions $x_P^{\mu}(\lambda)$ transform as 
\begin{equation}
 x_P^{\mu}(\lambda) \to x_P^{\mu}(\lambda) - \zeta(\lambda) e_P^{\mu}(\lambda), \label{worldlinerepar}
 \end{equation}
 where $e^{\mu}_P$ is defined as the first worldline derivative
 \begin{equation}
 \quad e_P^{\mu}(\lambda) \equiv \frac{d x_P^{\mu}(\lambda)}{d\lambda} \label{emu}.
 \end{equation}
The action thus transforms under worldline reparametrization \cref{worldlinerepar} as
\begin{equation}
 \delta S_P = - \int d\lambda \frac{\delta S_P}{\delta x_P^{\mu}(\lambda)}e_P^{\mu}(\lambda)\zeta(\lambda).
\end{equation}
By requiring this expression to vanish for arbitrary $\zeta(\lambda)$, one obtains
\begin{equation}
 \frac{\delta S_P}{\delta x_P^{\mu}(\lambda)}e_P^{\mu}(\lambda) = 0, \label{eqn:identity-reparametrization}
\end{equation}
as an identity holding even without considering the particle equation of motion.

\paragraph{Variations of the point particle action} For simplicity, we now assume that the point particle action depends on the particle position $x_P^{\mu}(\lambda)$ only through its first derivative $e_P^{\mu}(\lambda)\equiv dx_P^{\mu}(\lambda)/d\lambda$ so that the action is invariant under a constant shift of $x_P^{\mu}(\lambda)$ and that the corresponding equations of motion are second order. We also assume that the action does not depend on second or higher derivatives of the gravitational scalars $\varphi^a(x)$ and any derivatives of the metric $g_{\mu\nu}(x)$. In this case, the action is of the form
\begin{equation}
 S_P[x_P^{\mu}(\lambda), \varphi^a(x), g_{\mu\nu}(x)] 
  = \int d^4x \!\!\int d\lambda \,\, \mathcal{L}\Big(e_P^{\mu}(\lambda), \varphi^a(x), \partial_{\mu}\varphi^a(x), g_{\mu\nu}(x)\Big) \delta^4(x-x_P(\lambda)),
\end{equation}
so that one can easily compute its variations as follows:
\begin{eqnarray}
 \frac{\delta S_P}{\delta x_P^{\mu}(\lambda)} &=&
  -\frac{d}{d\lambda}\left( \left.\frac{\partial \mathcal{L}}{\partial e_P^{\mu}}\right|_{x_P(\lambda)}\right) + \left.\frac{\partial\mathcal{L}}{\partial\varphi^a}\partial_{\mu}\varphi^a\right|_{x_P(\lambda)} + \left.\frac{\partial\mathcal{L}}{\partial (\partial_{\nu}\varphi^a)}\partial_{\mu}\partial_{\nu}\varphi^a\right|_{x_P(\lambda)} + \left.\frac{\partial\mathcal{L}}{\partial g_{\mu\nu}}\partial_{\mu}g_{\mu\nu}\right|_{x_P(\lambda)},\nonumber\\
\mathcal{C}_a(x) &=& \frac{2}{\sqrt{-g}}\frac{\delta S_P}{\delta \varphi^a(x)} = 2 \int d\lambda \left[\frac{\partial\mathcal{L}}{\partial\varphi^a}  - \partial_{\mu}\left(\frac{\partial\mathcal{L}}{\partial (\partial_{\mu}\varphi^a)}\right)\right]\frac{\delta^4(x-x_P(\lambda))}{\sqrt{-g}}, \nonumber\\
T^{\mu\nu}(x) &=& \frac{2}{\sqrt{-g}}\frac{\delta S_P}{\delta g_{\mu\nu}(x)} = 2\int d\lambda \frac{\partial\mathcal{L}}{\partial g_{\mu\nu}}\frac{\delta^4(x-x_P(\lambda))}{\sqrt{-g}}. 
\end{eqnarray}
One can easily see that $\mathcal{C}_a$ and $T^{\mu\nu}$ are covariant. On the other hand, it is less obvious that $\delta S_P/\delta x_P^{\mu}(\lambda)$ is also covariant. In order to show the covariance of $\delta S_P/\delta x_P^{\mu}(\lambda)$  explicitly, we define the quantities
\begin{equation}
X^{ab}(x) \equiv g^{\mu\nu}(x)\partial_{\mu}\varphi^a(x)\partial_{\nu}\varphi^b(x), \quad
  Y^a(x,\lambda) \equiv e_P^{\mu}(\lambda)\partial_{\mu}\varphi^a(x), \quad
  Z(x,\lambda) \equiv g_{\mu\nu}(x)e_P^{\mu}(\lambda)e_P^{\nu}(\lambda), 
  \end{equation}
and use them to rewrite $\mathcal{L}$ as
\begin{equation}
 \mathcal{L} = \mathcal{L}(X^{ab}(x), Y^a(x,\lambda), Z(x,\lambda), \varphi^a(x)).
\end{equation}
It is then straighforward to show that
\begin{eqnarray}
 \frac{\delta S_P}{\delta x_P^{\mu}(\lambda)} &=& \frac{\partial\bar{\mathcal{L}}}{\partial\bar{X}^{ab}}\left.\partial_\mu X^{ab}\right|_{x_P(\lambda)} - \frac{d}{d\lambda}\left(\frac{\partial\bar{\mathcal{L}}}{\partial\bar{Y}^a}\right)\left.\partial_{\mu}\varphi^a\right|_{x_P(\lambda)}\nonumber\\
& & - 2\frac{d}{d\lambda}\left(\frac{\partial\bar{\mathcal{L}}}{\partial\bar{Z}}\right) e^P_{\mu} - 2\frac{\partial\bar{\mathcal{L}}}{\partial\bar{Z}} \frac{D e_{\mu}}{D\lambda} + \frac{\partial\bar{\mathcal{L}}}{\partial\bar{\varphi}^a}\left.\partial_{\mu}\varphi^a\right|_{x_P(\lambda)},
\label{pp-eom-covariant}
\end{eqnarray}
where the overlines denote quantities evaluated on the particle worldline, namely
\begin{align}
 &\bar{X}^{ab} \equiv X^{ab}(x_P(\lambda)), \quad \bar{Y}^a \equiv Y^a(x_P(\lambda),\lambda), \quad \bar{Z} \equiv Z(x_P(\lambda),\lambda), \quad \bar{\varphi}^a \equiv \varphi^a(x_P(\lambda)),\\
 &\bar{\mathcal{L}} \equiv \mathcal{L}(\bar{X}^{ab}, \bar{Y}^a, \bar{Z}, \bar{\varphi}^a).
\end{align}
Finally the down index vector  $e^P_{\mu}$ and its derivative are naturally expressed from $e_P^{\mu}$ of \cref{emu} through
\begin{equation}
 e^P_{\mu} \equiv g_{\mu\nu}(x_P(\lambda))e_P^{\nu}(\lambda), \quad
  \frac{D e^P_{\mu}}{D\lambda} \equiv   \frac{d e^P_{\mu}}{d\lambda} - \Gamma^{\rho}_{\ \mu\nu}e_P^{\nu}e^P_{\rho}, 
\end{equation}
where $\Gamma^{\rho}_{\ \mu\nu}$ are the space-time Christoffel symbols. The variation (\ref{pp-eom-covariant}) is now manifestly covariant.

\paragraph{Massive point particle} 

As a concrete point particle action that respects the diffeomorphism and reparametrization invariance, let us consider the following action with a field dependent mass and a Lorentz-type coupling:
\begin{equation}
S_P= \int d^4x \int d\lambda \left\{ - m(\varphi) \sqrt{-g_{\mu\nu}\frac{dx_P^{\mu}}{d\lambda}\frac{dx_P^{\nu}}{d\lambda}} +h_a(\varphi) \frac{dx_P^{\mu}}{d\lambda}\partial_{\mu}\varphi^a\right\}\delta^4(x-x_P(\lambda))d\lambda, \label{ppactionspecialcaseextension_mainBody}
\end{equation}
where $m(\varphi)$ and $h_a(\varphi)$ are functions of the gravitational scalars $\varphi=\{\varphi^a\}$. This corresponds to 
\begin{equation}
 \bar{\mathcal{L}} = -m(\bar{\varphi})\sqrt{-\bar{Z}} + h_a(\bar{\varphi})\bar{Y}^a,
\end{equation}
leading to the equations of motion for $x_P^{\mu}(\lambda)$ of the form 
\begin{equation}
 \frac{1}{\sqrt{-\bar{Z}}}\frac{D}{D\lambda}\left(\frac{m(\bar{\varphi})e^P_{\mu}}{\sqrt{-\bar{Z}}}\right) = F_{\mu}, \label{eom-pp-force-pre}
\end{equation}
where
\begin{equation}
 F_{\mu} \equiv \left[ - \frac{\partial m(\bar{\varphi})}{\partial\bar{\varphi}^a} + \left(\frac{\partial h_b(\bar{\varphi})}{\partial\bar{\varphi}^a}-\frac{\partial h_a(\bar{\varphi})}{\partial\bar{\varphi}^b}\right) \frac{\bar{Y}^b}{\sqrt{-\bar{Z}}}\right]\left.\partial_{\mu}\varphi^a\right|_{x_P(\lambda)}.
\end{equation}
This equation is a extended version of the one obtained in \cite{Eardley:1975,Nordtvedt:1968qs} in the case of a varying mass only. On defining the $4$-velocity $u^{\mu}$ of the particle and its acceleration $Du^{\mu}/D\tau$ as
\begin{equation}
 u^{\mu} \equiv \frac{e_P^{\mu}}{\sqrt{-\bar{Z}}} = \frac{d x_P^{\mu}}{d\tau}, \quad \frac{D u^{\mu}}{D\tau} \equiv \frac{d u^{\mu}}{d\tau} + u^{\rho}u^{\sigma}\Gamma^{\mu}_{\ \rho\sigma}|_{x_P(\tau)}, \quad d\tau = \sqrt{-\bar{Z}}d\lambda,
\end{equation}
and using the identity \eqref{eqn:identity-reparametrization}, the equation of motion for $x_P^{\mu}$ can be rewritten as
\begin{equation}
 m(\bar{\varphi}) \frac{D u^{\nu}}{D\tau} = F_{\mu} \gamma^{\mu\nu}|_{x_P(\tau)}, \quad
  F_{\mu} = \left[-\partial_a m + (\partial_a h_b - \partial_b h_a) u^{\alpha}\partial_{\alpha}\varphi^b\right] \partial_{\mu}\varphi^a|_{x_P(\tau)},  \label{eom-pp-force}
\end{equation}
where 
\begin{equation}
 \gamma^{\mu\nu} \equiv g^{\mu\nu} + u^{\mu}u^{\nu}
\end{equation}
projects spacetime indices onto the spatial section orthogonal to the worldline.

On the other hand, the variation of the action with respect to $\varphi^a$ gives
\begin{equation}
 \mathcal{C}_a = 2\int d\tau \left[ -\partial_a m + (\partial_a h_b - \partial_b h_a) u^{\alpha}\partial_{\alpha}\varphi^b \right] \frac{\delta^4(x-x_P(\tau))}{\sqrt{-g}}.  \label{Ca-pp}
\end{equation}
Using equation \eqref{DmuTmunu-pp} it implies that
\begin{equation}
\nabla^{\nu}T_{\mu\nu} =  \frac{1}{2}\mathcal{C}_a\partial_{\mu}\varphi^a = \int d\tau F_{\mu}(\lambda)\frac{\delta^4(x-x_P(\tau))}{\sqrt{-g}}.\label{divT-F-pp}
\end{equation}
By comparing \cref{eom-pp-force} or \cref{eom-pp-force-pre} to equation \eqref{divT-F-pp}, one can see that the divergence of the stress-energy tensor is directly related to the force applied on the point particle. Put another way, the direct couplings are responsible for forces expected to deviate the point particle from the metric geodesics. The next subsection is devoted to the study of the consequences of such forces.

\subsection{Massive body dynamics}\label{ppparticulardynamics}

As we have just seen in the previous subsection, the coupling $\mathcal{C}_a$ is related to the point particle equation of motion. As motivated in the introduction of the current section, the constraint on the $\beta$ parameter coming from massive test bodies dynamics can be sketched in the simplified Eddington-Robertson-Schiff (ERS) formalism. In this part we thus derive the massive body particle equation of motion at this level of approximation.

\paragraph{Order of the point-particle direct couplings} We start by expanding the couplings $\mathcal{C}_a$ appearing in the equation of motions, in the case of a point particle Lagrangian depending on gravitating scalars as in \cref{ppactionspecialcaseextension_mainBody}. Their expressions are given in \cref{Ca-pp}. With use of the PPN expansion for the different background fields $g_{\mu\nu}$ and $\varphi^a$ given from \cref{metricPPNexpansion_00,metricPPNexpansion_0i,metricPPNexpansion_jk,scalarPPNexpansion,scalarPPNexpansion2}. The couplings, and so the terms appearing in the equation of motion \cref{eom-pp-force}, are thus expanded as:
\begin{align}
\mathcal{C}_a &=
\frac{2}{\sqrt{-g}}\int d\tau \left\{-\partial_a m + \left(\partial_ah_b-\partial_bh_a\right)u^{\alpha}\partial_{\alpha}\varphi^b\right\}\delta^{(4)}(x-x_P(\tau))\nonumber\\
&=\frac{2}{\sqrt{-g}} \int d\tau \left\{ -\partial_a m(\varphi_0^b) - \partial_{ab} m(\varphi_0^c) \varphi_2^b + \big(\partial_ah_b(\varphi_0^c)-\partial_bh_a(\varphi_0^c)\big) u^{\alpha} \partial_{\alpha}\varphi_2^b \right\}  \nonumber\\
&\hspace{2.5cm} \times \delta^{(4)}\big(x-x_P(\tau)\big) +   O(\epsilon^4). \label{couplingexpansion}
\end{align}
We see that, at the lowest order, this amounts to take a  coupling of the form:
\begin{equation}
\mathcal{C}_a= -\frac{2}{\sqrt{-g}}\int \partial_a m(\varphi_0^b) \delta^{(4)}(x-x_P(\tau))d\tau + O(\epsilon^2). 
\end{equation}
When this coupling is not vanishing, one can safely ignore the Lorentz-type coupling $h_a$ in \cref{ppactionspecialcaseextension_mainBody}. Nevertheless, when this dominant coupling vanishes, namely for non-varying mass $\partial_am\equiv0$, the Lorentz-type couplings provides the leading contribution to $\mathcal{C}_a$. 

We could expand the point particle coupling obtained above in the same way as for the source, given by \cref{ccaexpansion}. In the point particle case at the lowest order, it simply reads
\begin{equation}
\mathcal{C}^a= - \frac{2}{\sqrt{-g}}\int c^a m \delta^4(x-x_P(\tau)) \, d\tau + O(\epsilon^2), \label{caPP}
\end{equation}
with the trivial identification
\begin{equation}
 c^a= -2 s^a  +O(\epsilon^2), \label{particularcouplings}
 \end{equation}
 where we defined the ``sensitivity'' in a way similar to \cite{Eardley:1975}:
\begin{equation}
s_a\equiv \partial_a\ln m(\varphi_0). \label{sensitivity}
\end{equation}

\paragraph{Equation of dynamics}
At the lowest non-trivial PPN order, the point particle equation of motion of \cref{eom-pp-force} reduces to:
\begin{align}
\frac{d^2x_P^{\mu}}{d\tau^2}+\Gamma^{\mu}_{\sigma\rho}\frac{dx_P^{\sigma}}{d\tau}\frac{dx_P^{\rho}}{d\tau}&= - \left[ \partial^{\nu} \varphi^a \partial_a \ln m + O(\epsilon^4) \right] \gamma^{\mu}_{\ \nu} |_{x_P(\tau)} \nonumber\\
&=-  2s_a\gamma_{\varphi^a} \left[ U^{,\nu} +O(\epsilon^4) \right] \gamma^{\mu}_{\ \nu} |_{x_P(\tau)}, \label{eomparticularexpandedPPN}
\end{align}
where in the last line we expanded the scalars according to \cref{scalarPPNexpansion}. The time component of \cref{eomparticularexpandedPPN}  reads:
\begin{equation}
\frac{dt^2}{d\tau^2}+\Gamma^{0}_{\sigma\rho}\frac{dx_P^{\sigma}}{d\tau}\frac{dx_P^{\rho}}{d\tau}=-  2s_a\gamma_{\varphi^a} U_{,j} \frac{dx_P^j}{d\tau}  \frac{dt}{d\tau}+O(\epsilon^4), \label{timecomp}
\end{equation}
where $\gamma^{\mu}_{\ \nu}$ has projected out the seemingly leading term containing $U^{,0}$. To study the spatial components of \cref{eomparticularexpandedPPN} we can switch from $\tau$ derivatives to $t$ derivatives by decomposing
\begin{equation}
\frac{d^2x_P^i}{d\tau^2}=\frac{dx_P^i}{dt}\frac{dt^2}{d\tau^2}+\frac{d^2x_P^i}{dt^2}\left(\frac{dt}{d\tau}\right)^2.
\end{equation}
Hence we obtain the following equality:
\begin{align}
&-\Gamma^{i}_{\sigma\rho}\frac{dx_P^{\sigma}}{d\tau}\frac{dx_P^{\rho}}{d\tau}- 2s_a\gamma_{\varphi^a} \left(U^{,i}+ U_{,\alpha} \frac{dx_P^\alpha}{d\tau}  \frac{dx_P^i}{d\tau}\right)
 \nonumber\\
& \qquad \qquad \qquad
=\frac{dx_P^i}{dt}\left(-\Gamma^{0}_{\sigma\rho}\frac{dx_P^{\sigma}}{d\tau}\frac{dx_P^{\rho}}{d\tau}-  2s_a\gamma_{\varphi^a} U_{,j} \frac{dx_P^j}{d\tau}  \frac{dt}{d\tau}\right)+\frac{d^2x_P^i}{dt^2}\left(\frac{dt}{d\tau}\right)^2,
\end{align}
which after multiplying by $\left(\frac{d\tau}{dt}\right)^2$ simplifies to
\begin{equation}
\frac{d^2x_P^i}{dt^2}+\left(\Gamma^{i}_{\sigma\rho}-\Gamma^{0}_{\sigma\rho}\frac{dx_P^i}{dt}\right)\frac{dx_P^{\sigma}}{dt}\frac{dx_P^{\rho}}{dt}=-2s_a\gamma_{\varphi^a} U^{,i}\left(\frac{dt}{d\tau}\right)^2+O(\epsilon^3).
\end{equation}
We see that this is the leading term responsible for deviation from geodesic motion. As we comment later, this deviation nevertheless happens already at Newtonian order. Using the expressions given in \cref{appendix:Christoffels} for the Christoffel symbols, where we put the $\Phi_i, V_i, W_i$ potentials to zero as shall be the case in the ERS formalism, we obtain the following corrected massive particle equation of motion: 
\begin{equation}
\frac{d^2x_P^i}{dt^2}=U_{,i}\left(\vphantom{1^1}1-2s_a \gamma_{\varphi^a}\right)+\gamma U_{,i}v^2+2(\gamma+\beta)U_{,i} U+ 2v^i(1+\gamma)\vec{v}\cdot \vec{\nabla}U+O(\epsilon^5). \label{eomparticlePPNexpanded2}
\end{equation}  
 As explained in the intermediate steps above, in the final result \cref{eomparticlePPNexpanded2} we only kept the leading order for the terms related to the direct couplings. These terms appear at Newtonian order, hence modifying the Newton laws of dynamics. This is not a surprise, and the above equation can be used to constrain the values of the direct coupling. Indeed, as it is independent of the velocity of the test body $\vec{v}$, it should satisfy $s_a \gamma_{\varphi^a}\ll v^2$ for typical velocities of test bodies in our solar system, otherwise it would be directly observable (and ruled out) by comparing the motion of various planets. In fact, if not small enough, such couplings to scalars (axions) should also be detectable by experiments on Earth.

 \paragraph{Lorentz-type coupling} When the probe particle mass does not depend on the scalar fields, $i.e.$ for vanishing sensitivity \eqref{sensitivity}, one should consider a coupling of the Lorentz-type form. This is the second term of \cref{ppactionspecialcaseextension_mainBody}, and its $\epsilon$ expansion is given in the second line of \cref{couplingexpansion}. As it contains $u^{\alpha}\partial_{\alpha}\varphi_2$, it is of order $\epsilon^3$. Hence this type of coupling does not play a role at the first PPN level of approximation, and the dynamics will thus be described by \cref{eomparticlePPNexpanded2} taking $s_a=0$, which are the standard PPN equations of motion without the direct couplings. 
 
\subsection{Photons dynamics and experiments on light rays}\label{photonsdynamics}

In this subsection, complementary to the previous ones, we study the dynamics of a massless relativistic particle. We study test photons dynamics and choose a particular case for the direct coupling of the gravitating scalars to photons. We indeed consider the simplest kind of direct couplings of photons with an axionic gravitating field $a(x^{\mu})$, which is expressed as follows
\begin{equation}
\frac{1}{\sqrt{-g}}\mathcal{L}_{m-a}= i c_2 \frac{\alpha}{f_a} a(x^{\mu}) \, F_{\mu\nu} \tilde F^{\mu\nu}(x^{\mu}). \label{couplingsaxion}
\end{equation}
We defined the dual electromagnetic tensor $\tilde F^{\mu\nu}=i/(2\sqrt{-g}) \epsilon^{\mu\nu\alpha\beta}F_{\alpha\beta}$, the fine structure constant $\alpha$ and the axion decay constant $f_a$. The $\mathcal{C}^a$ functions can in principle be extracted from the above Lagrangian and the one describing the coupling of the axion to the electrons and nucleons constituting the matter sources. This would require a complete description of the source from the microscopic scales up to the macroscopic scales and would thus be beyond the scope of the present work. In this subsection we shall simply show that the coupling \cref{couplingsaxion} does not change the standard PPN constraints from test photons. 

The equations of motion derived from the lagrangian \eqref{couplingsaxion} are the modified Maxwell equations:
\begin{equation}
\nabla_\nu F^{\mu\nu}=\frac{i\alpha c_2}{f_a}  \tilde F^{\mu\nu} \nabla_{\nu}a = - \frac{\alpha c_2}{f_a}  \frac{\epsilon^{\mu\nu\rho\sigma}}{\sqrt{-g}} F_{\rho\sigma} \nabla_{\nu}a . 
\end{equation}
In the presence of a (static) axion source, this term is not vanishing and is proportional to the gradient of the axion profile. According to \eqref{scalarPPNexpansion}, the axion field is given by 
\begin{equation}
a(x^{\mu})=a_0 + 2\gamma_a U(x^{\mu}) + O(\epsilon^4).
\end{equation}

When considered together with the Lorenz gauge condition $\nabla_{\mu} A^{\mu}=0$ and the definition $F_{\mu\nu}\equiv\nabla_{\mu}A_{\nu}-\nabla_{\mu}A_{\nu}$, the above field equation is written as
\begin{equation}
\nabla^{\mu}\nabla_{\mu}A^{\nu}+R^{\nu}_{\,\,\mu}A^{\mu}=- \frac{\alpha c_2}{f_a}  \frac{\epsilon^{\nu\mu\rho\sigma}}{\sqrt{-g}} \nabla_{\mu}A_{\sigma} \nabla_{\rho}a. \label{photonEM}
\end{equation}
where $R_{\mu\nu}$ is the space-time Ricci curvature. Far from the source, the space-time geometry characteristic length $L_g$ is very large compared to the wavelengths $\lambda$ of typical detectable photons, $i.e.$ $L_g\gg \lambda$. If we furthermore assume that the wave-packet characteristic length $L_w$ is large in front of the wavelength  $\lambda$, we can use the WKB approximation to solve \cref{photonEM}. Namely, we decompose the gauge potential as
\begin{equation}
A_{\mu}\equiv\mathcal{A}_{\mu}e^{i\phi}\equiv \mathcal{A}e_{\mu}e^{i\phi} , \qquad \mathcal{A}\equiv \sqrt{-\mathcal{A}^{\mu}\mathcal{A}_{\mu}}, \,\, e^{\mu}e_{\mu}=-1, 
\end{equation}
and suppose that the various quantities vary as:
\begin{align}
  &k_{\mu}\equiv \partial_{\mu} \phi \sim \lambda^{-1}, \qquad \nabla_{\nu} k_{\mu} \sim (\lambda L_w)^{-1},\nonumber\\
  &\mathcal{A}_{\mu} \sim O(1), \qquad \nabla_{\nu} \mathcal{A}_{\mu} \sim {L_w}^{-1},\\
  &\nabla_{\rho}a \sim {L_g}^{-1}. \nonumber
  \end{align}
  The length scale $L_g$ can be extracted from the Riemann curvature tensor and is typically of order $L_g\sim c r^{3/2}/\sqrt{GM}$, where $r$ is the distance between the matter source and the place where the photons propagate. The Riemann curvature could in general be extracted from the PPN metric, but as long as the PPN parameters stay small, the length $L_g$ is close to its general relativity value. Hence, the dominant contribution in $\lambda^{-1}$ in the WKB approximation of \cref{photonEM} is simply:
  \begin{equation}
  -k^{\mu}k_{\mu}\mathcal{A}^{\nu}=0, \qquad O(\lambda^{-2}),
  \end{equation}
  which leads to $k^{\mu}k_{\mu}=0$ and to the standard geodesic equation $k^{\mu}\nabla_{\nu}k_{\mu}=0$ for photons. The Lorenz gauge condition also leads to:
  \begin{equation}
  k_{\mu}\mathcal{A}^{\nu}=0, \qquad O(\lambda^{-1}),
  \end{equation}
  showing that the gauge potential is orthogonal to the direction of propagation. On the other hand, the second dominant contribution in the WKB approximation of \cref{photonEM} reads:
  \begin{equation}
\mathcal{A}^{\mu}\nabla_{\nu}k^{\nu} + 2 k^{\mu}\nabla_{\mu}\mathcal{A}^{\nu} + \frac{ 2}{\sqrt{-g}} \mathcal{A}_{\beta}k_{\gamma} \nabla_{\alpha} a \epsilon^{\mu\alpha\beta\gamma}=0, \qquad O(\lambda^{-1}).
\end{equation}   
It can be solved by:
\begin{align}
k^{\nu}\nabla_{\nu}e^{\mu}=0, \qquad e^{\mu}\nabla_{\alpha}(\mathcal{A}^2k^{\alpha})=-\frac{2}{\sqrt{-g}}\epsilon^{\mu\alpha\beta\gamma}\mathcal{A}^2e_{\beta}k_{\gamma}\nabla_{\alpha} a.
\end{align}
We see that in the WKB approximation, test photons follow geodesics of the PPN metric but their polarization and amplitude  are sensitive to the presence of the axionic background. This last aspect is at the center of axion searches through direct conversion or change in polarization of light in the presence of magnetic fields \cite{Sikivie:1983ip,Sikivie:2006ni,Sikivie:2020zpn}. On the other hand, as the photons follow the PPN metric, classical experiments measuring deviation or time delays of light will directly probe the PPN metric and give access to the $\gamma$ parameter \cite{Will:1993ns,Lambert:2009xy,Will:2014kxa}.

\section{Conclusions}\label{section:conclusions}

In this work, we studied the implications of adding direct couplings between gravitating scalar fields and matter, on top of the universal coupling to the metric, in multi-scalar tensor theories of gravity. Such direct couplings are expected to have direct effects since they modify the way space-time geometry is influenced by a localized matter source. This observation motivated the study of the weak gravity quasi-static regime of such theories, which is the relevant regime to describe solar-system tests of gravity. A central question addressed in this work was related to the possibility of screening cosmologically active scalars through their direct couplings to matter. Such screening would have made their presence undetectable in solar systems experiments.

In the weak gravity quasi-static regime, theories of gravity can be compared with each other through the parametrized post-Newtonian formalism. We used this formalism and computed the complete expression for the ten PPN parameters in multi-scalar theories of gravity with direct couplings. The expressions we derived allow to evaluate the PPN parameters for the considered class of scalar-tensor theories of gravity. The latter are theories including several massless gravitating scalars with a curved scalar target-space and coupled non-universally to the matter sector. We showed that the $\zeta_3$ and $\zeta_4$ PPN parameters are indeed modified by the presence of the direct couplings, even after eliminating the direct couplings of the gravitating scalars to the energy density ($c^a$ in \eqref{ccaexpansion}) of the source by redefinition of the Jordan frame. On the other hand, all other PPN parameters including $\beta$ and $\gamma$ are the same as the ones obtained in the same multi-scalar theory without direct couplings, provided that the Jordan frame is properly identified. This conclusion seems to go against the screening mechanism through direct couplings invoked by recent works in the literature.

We then studied if such couplings, even small, would change the classical tests of gravity in the weak-field regime, by modifications in the way probes would move on the PPN background. To make our arguments as general as possible, we did not assume {\it a priori}  that the direct couplings are the same for the probe and the source of gravity. Therefore the coefficient $c^a$ in the expansion \eqref{ccaexpansion} of the direct couplings may remain non-vanishing for the probe even after setting $c^a=0$ for the source by redefinition of the Jordan frame. As expected, direct couplings for the probe are responsible for additional direct forces on top of the gravitational force mediated through the space-time geometry. Hence, large direct couplings would be directly observable and would be ruled out. We supported this intuition by studying the dynamics of a massive point particle directly coupled to gravitating scalars, in presence of metric and scalar backgrounds generated by a PN matter source. For the point particle couplings under considerations, we deduced that if the couplings are small enough not to perturb the Newtonian order, test particles will follow the PPN metric geodesics, giving access to the $\gamma$ and $\beta$ PPN parameters of the theory. The classical constraints will thus apply identically to theories with or without direct couplings, as far as the direct couplings of the probe do not spoil the success of Newton gravity in its regime of applicability.  Experiments involving photons in the regime of validity of the WKB approximation would also give access to the $\gamma$ parameter, as in the standard PPN formalism.

Possible extensions of this work would include the complete study  in the PPN formalism of the two-body system, made of a PN source (Sun) and a massive probe (a planet), in theories with direct couplings. This should include the study of (non-)conservation laws for the different PN quantities (densities, momentum...) for the massive bodies, used when integrated the fluid equations of motion in the interior of massive bodies. Although technically involved, the procedure is well defined. The result would allow to give quantitative predictions for the results of solar system experiments such as the measure of the perihelion shift of Mercury.

\acknowledgments
This article is based upon work from COST Action COSMIC WISPers CA21106, 
supported by COST (European Cooperation in Science and Technology). The work of OL was supported in part by Japan Society for the Promotion of Science Grant-in-Aid for Scientific Research No. 17H06359. The work of SM was supported in part by Japan Society for the Promotion of Science Grants-in-Aid for Scientific Research No. 17H02890, No. 17H06359, and by World Premier International Research Center Initiative, MEXT, Japan.

\appendix

\section{Notations and identities for multi-scalar gravity and PPN formalism}\label{appendix:usefulrelations}
\subsection{Scalar target-space functions} \label{appendix:Targetspacefunctions}
The multi-scalar theory considered in this paper is based on the action \eqref{Action}. To give an explicit expression of the gravitating sector, one should specify the non-minimal coupling function $F(\varphi^a)$ and the target-space metric $\mathcal{G}_{ab}(\varphi^c)$. As shown in the main body of the paper, the field equations and their PN expansions are easily expressed in terms of the following functions:
\begin{align}
&\Gamma^a_{bc}=\frac 12 \Gc^{ad}(\p_b\Gc_{cd}+\p_c\Gc_{bd}-\p_d\Gc_{bc}), \quad F_a\equiv \partial_a F, \quad F_{ab}=\partial_{a}\partial_{b}F,\\
 &A^a_b\equiv\frac{F^aF_b}{F}=B^aF_b, \quad B^a\equiv\frac{F^a}{F}, \qquad C^a_{bc}=\Gamma^a_{bc}+\frac{B^a}2(\Gc_{bc}+3F_{bc}),\\
&\tilde F_{cd}\equiv\nabla_cF_d= \nabla_c\partial_dF=F_{cd}-\Gamma^e_{cd}F_e.
\end{align}
One can use different sets of variables, such as the one naturally appearing when studying the Lagrangian $\{ F, F_a, F_{ab}, \mathcal{G}_{ab}, \Gamma^a_{bc}\}$, or the one related to the expansions of the field equations $\{ F, B_a, A^a_b, F_{ab}, \mathcal{G}_{ab}, C^a_{bc}\}$, or yet the naturally target-space covariant one $\{B_a,F_a,\tilde{F}_{ab},\mathcal{G}_{ab},\Gamma^a_{bc}\}$. These sets are only used for convenience and do not constitute bases of target-space functions. We give a couple of examples of relations used to go from one set to another: 
\begin{align}
B^cF_d\,\partial_cB^d+B^cB^dF_{cd}&=-\frac{B^cB^d}{F}(F_cF_d-2F\tilde{F}_{cd}),\\
2 B^cB^dF_e C^e_{cd}&=  B^cB^dF_{cd}(3B^eF_e+1)- B^cF_d \partial_cB^d\nonumber\\
&=B^cB^dF_{cd}(3B^eF_e+2)+\frac{B^cB^d}{F}(F_cF_d-2F\tilde{F}_{cd}).
\end{align}

\subsection{Parametrized post-Newtonian functionals}  \label{appendix:PNPotentials}

In terms of the source rest-mass density $\rho$, the Newtonian potential is given by
\begin{equation}
U(\vec{x},t)\equiv\int \frac{\rho(\vec{x}',t)}{|\vec{x}-\vec{x}'|}d^3x'.
\end{equation}
The post-Newtonian functionals and potentials are defined as follows:
\begin{align}
&V_i\equiv\int\frac{\rho v_i(\vec{x}',t)}{|\vec{x}-\vec{x}'|}d^3x', \qquad W_i\equiv\int\frac{\rho \vec{v}(\vec{x}',t)\cdot(\vec{x}-\vec{x}')(x-x')_i}{|\vec{x}-\vec{x}'|^3}d^3x',\\
&\Phi_W\equiv\int \rho(\vec{x'},t)\rho(\vec{x}'',t)\frac{\vec{x}-\vec{x}'}{|\vec{x}-\vec{x}'|^3}\cdot \left(\frac{\vec{x}'-\vec{x}''}{|\vec{x}'-\vec{x}''|}-\frac{\vec{x}-\vec{x}''}{|\vec{x}-\vec{x}''|}\right)d^3x'd^3x'',\\
&\Phi_1=\int \frac{\rho v^2(\vec{x}',t)}{|\vec{x}-\vec{x}'|}d^3x', \quad \Phi_2=\int \frac{\rho U (\vec{x}',t)}{|\vec{x}-\vec{x}'|}d^3x', \quad \Phi_3=\int \frac{\rho \Pi (\vec{x}',t)}{|\vec{x}-\vec{x}'|}d^3x', \quad \Phi_4=\int \frac{p(\vec{x}',t)}{|\vec{x}-\vec{x}'|}d^3x' \nonumber \\
&\mathcal{A}\equiv\int \frac{\rho(\vec{x}',t) \left[\vphantom{1^{1^1}}\vec{v}(\vec{x}',t)\cdot (\vec{x}-\vec{x}')\right]^2}{|\vec{x}-\vec{x}'|^3}d^3x', \quad \mathcal{B}\equiv\int \frac{\rho(\vec{x}',t)}{|\vec{x}-\vec{x}'|}  (\vec{x}-\vec{x}')\cdot \frac{d \vec{v}}{dt}(\vec{x}',t)  d^3x'
\end{align}
By using the Newtonian order Euler continuity equation
\begin{equation}
\frac{\partial \rho}{\partial t} + \vec{\nabla}(\rho \vec{v})=0,
\end{equation}
leading to the relation
\begin{equation}
\frac{\partial}{\partial t} \int \rho(\vec{x}',t) f(\vec{x},\vec{x}')d^3x'= \int \rho \vec{v} (\vec{x}',t) \cdot \vec{\nabla} f(\vec{x},\vec{x}')d^3x',
\end{equation}
one obtains, in addition to the standard Poisson equation for the Newtonian potential 
\begin{equation}
\nabla^2 U=-4\pi \rho,
\end{equation} the following useful relations:
\begin{align}
&\nabla^2 V_i=-4\pi \rho v_i, \quad \frac{\partial V_{i}}{\partial x^i}=-\frac{\partial U}{\partial t}, \\
&\nabla^2 \Phi_1=-4\pi \rho v^2, \quad \nabla^2 \Phi_1=-4\pi \rho U, \quad \nabla^2 \Phi_1=-4\pi \rho \Pi, \quad \nabla^2 \Phi_4=-4\pi p.
\end{align}
In the intermediate steps, one often uses the additional functionals,
\begin{equation}
\chi\equiv- \int {\rho(\vec{x}',t)}{|\vec{x}-\vec{x}'|}d^3x', \qquad U_{ij}\equiv\int\frac{\rho(\vec{x}',t) (x-x')_i (x-x')_j}{|\vec{x}-\vec{x}'|^3}d^3x',
\end{equation}
satisfying the following relations
\begin{equation}
\chi_{,ij}=-\delta_{ij} U + U_{jk}, \qquad \chi_{,0i}=V_i-W_i, \qquad \chi_{,00}=\mathcal{A}+\mathcal{B}-\Phi_1.
\end{equation}

\subsection{Christoffels symbols for the PPN metric}\label{appendix:Christoffels}

The Christoffel symbols for the metric in the standard PPN gauge, in the specific case where $\xi=\alpha_1=\alpha_2=\alpha_3=\alpha_4=\zeta_1=0$, take the following forms:
\begin{align}
&\Gamma^0_{00}=- U_{,0}\,,\qquad \Gamma^0_{0i}=- U_{,i}\,,  \\
& \Gamma^0_{ij}=\gamma \delta_{ij}U_{,0}+\left(2\gamma+\frac{3}{2}\right)V_{(i,j)}+\frac{1}{2}W_{(i,j)}\,, \\
&\Gamma^i_{0j}=\gamma \delta_{ij} U_{,0} -(2\gamma+2)V_{[i,j]}\,, \\
& \Gamma^i_{jk}=\gamma \left(\delta_{ij} U_{,k}+\delta_{ik} U_{,j}-\delta_{jk} U_{,i}\right),\\
&\Gamma^i_{00}=-U_{,i}+\frac{\partial}{\partial x^i}\left[(\beta+\gamma)U^2-\Phi\right ]-\frac{\partial}{\partial t} \left[(2\gamma+\frac{3}{2})V_{i}+\frac{1}{2} W_i \right],
\end{align}
where here $\Phi$ is defined as the combination
\begin{align}
&\Phi\equiv(\gamma+1)\Phi_1+(3\gamma-2\beta+1+\zeta_2)\Phi_2+(1+\zeta_3)\Phi_3+3(\gamma+\zeta_4)\Phi_4.
\end{align}

\bibliography{PPN}
\bibliographystyle{JHEP}

\end{document}